\documentclass[12pt,a4wide]{article}
\usepackage{afterpage}
\usepackage{epsfig}
\usepackage[a4paper]{geometry}
\usepackage{float}
\usepackage[stable]{footmisc}
\usepackage[utf8x]{inputenc}
\usepackage{a4wide}
\usepackage{amsmath,amssymb,amstext,amsthm,amssymb}
\usepackage{amsfonts}
\usepackage{framed}
\usepackage{graphicx}
\usepackage{bbold}
\usepackage{slashed} 
\usepackage{tensor} 
\usepackage{braket} 
\usepackage{array}
\usepackage{graphicx}
\usepackage{setspace}
\usepackage{hyperref}
\usepackage{overpic}
\usepackage{bbm}
\usepackage{cite}
\usepackage{color}
\usepackage{lipsum}
\usepackage{bm}

\DeclareSymbolFontAlphabet{\mathbb}{AMSb}

\newcommand{\del}[0]{\partial}

\let\baraccent=\=
\renewcommand{\=}[1]{\stackrel{#1}{=}}
\newcommand{\id}[0]{\mathbb{I}}

\newcommand{\eps}[0]{\varepsilon}

\newcommand{\zc}{z_{\text{cf}}}

\begin{document}

\pagestyle{plain}

\makeatletter
\@addtoreset{equation}{section}
\makeatother
\renewcommand{\theequation}{\thesection.\arabic{equation}}
\pagestyle{empty}

\vspace{0.5cm}

\begin{center}
{\Large \bf{Conifold Vacua with Small Flux Superpotential}
		\\[15mm]}
	\normalsize{Mehmet Demirtas, Manki Kim, Liam McAllister, and Jakob Moritz  \\[6mm]}
	\small{\slshape
		Department of Physics, Cornell University, Ithaca, NY 14853, USA
		 \\[8mm]  }
	\normalsize{\bf Abstract} \\[8mm]
\end{center}
\begin{center}
	\begin{minipage}[h]{15.0cm}
		
We introduce a method for finding flux vacua of type IIB string theory in which the flux superpotential is exponentially small and at the same time one or more complex structure moduli are stabilized exponentially near to conifold points.

	\end{minipage}
\end{center}
\newpage
\setcounter{page}{1}
\pagestyle{plain}
\renewcommand{\thefootnote}{\arabic{footnote}}
\setcounter{footnote}{0}

\tableofcontents

\newpage

\section{Introduction}

In order to understand the cosmologies that are possible in quantum gravity, one can search for de Sitter solutions in compactifications of string theory.  Kachru, Kallosh, Linde, and Trivedi (KKLT) famously proposed that orientifold compactifications of type IIB string theory that contain specific `components' in the right proportions will admit parametrically controlled de Sitter vacua \cite{Kachru:2003aw}.

These components --- a small classical flux superpotential \cite{Ashok:2003gk,Denef:2004ze,Denef:2004dm,Demirtas:2019sip}, a warped throat region \cite{Klebanov:2000hb,Giddings:2001yu,Giddings:2005ff,Douglas:2007tu,Blumenhagen:2016bfp,Martucci:2016pzt}, a potential for the K\"ahler moduli from Euclidean D3-branes or strong gauge dynamics \cite{Witten:1996bn,Koerber:2007xk,Blumenhagen:2009qh,Grimm:2011dj,Bianchi:2011qh,Bianchi:2012kt,Hamada:2018qef,Hamada:2019ack,Carta:2019rhx,Gautason:2019jwq,Bena:2019mte,Kachru:2019dvo}, and a supersymmetry-breaking sector from anti-D3-branes \cite{Kachru:2002gs,Bena:2009xk,Dymarsky:2011pm,Bena:2018fqc,Armas:2018rsy,Blumenhagen:2019qcg,Randall:2019ent,Dudas:2019pls} --- are by now rather well understood \emph{separately}. A remaining challenge in the pursuit of explicit examples of KKLT de Sitter vacua is to exhibit Calabi-Yau orientifolds that contain all these components at once, through calculations in which
corrections to the leading approximations are demonstrably well-controlled.

In this work we present a method for finding flux vacua that contain \emph{both} a warped throat region and an exponentially small classical flux superpotential, $|W_0| \ll 1$.
We do so by building on our recent work \cite{Demirtas:2019sip}, where we showed how to find flux vacua with $|W_0| \ll 1$.
In \cite{Demirtas:2019sip} we took all complex structure moduli to be near large complex structure (LCS).
Warped throats, on the other hand, occur in flux vacua in which one or more complex structure moduli are stabilized near conifold singularities --- we refer to such vacua as \emph{conifold vacua}.

If the quantized fluxes threading the A-cycle and B-cycle of a conifold in such a vacuum are sufficiently large, then the conifold region is accurately described by the warped deformed conifold supergravity solution found by Klebanov and Strassler \cite{Klebanov:2000hb}, and can serve as a setting for metastable supersymmetry breaking by anti-D3-branes \cite{Kachru:2002gs}. In the opposite regime of small 't Hooft coupling, the conifold region is accurately described by a cascading gauge theory that potentially has a metastable supersymmetry-breaking state --- gauge theory vacua of this sort have been analyzed in \cite{Argurio:2006ny,Retolaza:2015nvh,Argurio:2019eqb,Argurio:2020dkg,Argurio:2020npm}.

In order to generalize the mechanism of \cite{Demirtas:2019sip} to include conifolds
one has to overcome the following obstacle. Introducing fluxes on the conifold cycles generates a conifold superpotential $W_{\text{cf}}$ that by itself cannot be tuned to be small. Thus, the total flux superpotential will be small in string units only if the large conifold superpotential is efficiently canceled by a comparably large contribution $W_{\text{bulk}}$ generated by fluxes on other cycles, i.e.~if
\begin{equation}\label{eq:fine-tuning}
|W_0| := |\langle W_{\text{flux}}\rangle|=|\langle W_{\text{cf}}\rangle+\langle W_{\text{bulk}}\rangle|\ll 1\, .
\end{equation}
To achieve such a cancellation, one must first accurately compute the conifold superpotential $W_{\text{cf}}$ in the vicinity of a conifold singularity.

In the first part of this work, we compute $W_{\text{cf}}$ analytically
in the case where the shrinking $S^3$ of the conifold is mirror dual to a shrinking curve.  In such a case $W_{\text{cf}}$ is obtained by resumming the instanton corrections from string worldsheets wrapping the shrinking curve in the mirror threefold.  We then show, along the same lines as \cite{Demirtas:2019sip}, that one can choose quantized fluxes leading to an exponentially precise cancellation of the form shown in eq.~\eqref{eq:fine-tuning}.  Finally, we present explicit examples of conifold vacua in an O3/O7 orientifold of a Calabi-Yau hypersurface $\tilde{X}$ with $h^{1,1}(\tilde{X})=99$ and $h^{2,1}(\tilde{X})=3$.

Let us be clear in advance about the scope of this work.  We will present a mechanism for constructing classical flux vacua in which $|W_0| \ll 1$ and at least one complex structure modulus is stabilized near a conifold, and we will illustrate the mechanism with
flux vacua of an explicit orientifold.
Although a long-term goal is to combine such results with K\"ahler moduli stabilization and a metastable uplift to a de Sitter solution,\footnote{See e.g.~\cite{Louis:2012nb} for an analysis of a de Sitter solution arising from an explicit flux vacuum.} we will not take these latter steps here.  Exhibiting ensembles of flux vacua that can be lifted to metastable de Sitter solutions is an ambitious task for the future.

The organization of this paper is as follows.
In \S\ref{sec:setup} we set our notation (\S\ref{ss:setup}), recall a few results about the large complex structure limit in Calabi-Yau moduli space (\S\ref{ss:lcs}), and review the mechanism of \cite{Demirtas:2019sip} for constructing vacua with small flux superpotential (\S\ref{ss:sfs}).
In \S\ref{sec:result} we present a mechanism for constructing conifold vacua with small flux superpotential.
To illustrate this, in \S\ref{sec:example} we examine a Calabi-Yau threefold $X$ with $h^{1,1}(X)=3$ and $h^{2,1}(X)=99$ (\S\ref{ss:x}); construct its mirror $\tilde{X}$, and an orientifold thereof (\S\ref{ss:tildex}); and exhibit conifold vacua with $|W_0| \ll 1$ in the orientifold of $\tilde{X}$ (\S\ref{sec:explicit}).
We conclude in \S\ref{sec:conclusions}.
The appendix contains two independent computations of the D3-brane tadpole in the orientifold of \S\ref{ss:tildex}.

\textbf{Note added:} Simultaneously with ours, the work \cite{Alvarez-Garcia:2020pxd} appeared which addresses the same issue.
\section{Vacua with small flux superpotential}\label{sec:setup}
\subsection{Setup}\label{ss:setup}

We will work in the landscape of four-dimensional $\mathcal{N}=1$ supergravity solutions obtained from compactifications of type IIB string theory on O3/O7 orientifolds of Calabi-Yau threefolds. While we are ultimately interested in analyzing the full vacuum structure of such models, arising in particular from the non-perturbative potential for the K\"ahler moduli, in this paper we will neglect the K\"ahler moduli altogether. That is, we consider the classical no-scale solutions of \cite{Giddings:2001yu}. Throughout this paper, unless noted otherwise, we will work in ten-dimensional Einstein frame in units where $\ell_s^2\equiv (2\pi)^2\alpha'=1$, and use our freedom to Weyl-rescale the four-dimensional metric to set the four-dimensional reduced Planck mass to one. These conventions match those of \cite{Demirtas:2019sip}.

To begin, we consider a Calabi-Yau threefold $\tilde{X}$ and a holomorphic and isometric involution $\tilde{\mathcal{I}}:\tilde{X}\rightarrow \tilde{X}$, with induced action on the holomorphic three-form $\Omega\mapsto -\Omega$. After the orientifolding, the fixed locus of $\tilde{\mathcal{I}}$ hosts O3-planes and O7-planes.

Let $Q$ be the total D3-brane charge of the O3-planes and seven-brane stacks. If $Q < 0$
then its contribution to the D3-brane tadpole can be canceled by $N_{D3}$ mobile D3-branes as well as three-form fluxes $F_3$ and $H_3$ such that\footnote{In our conventions a mobile D3-brane has a single unit of D3-brane charge, while a D3-brane frozen onto the orientifold fixed locus has charge $1/2$.}
\begin{equation}
Q_{D3}^{\text{total}}=N_{D3}+\frac{1}{2}\int_{\tilde{X}} F_3\wedge H_3+Q=0\, .
\end{equation}
Let $\{\Sigma_{(3)a},\Sigma_{(3)}^a\}$ be a symplectic basis of $H_3(\tilde{X},\mathbb{Z})$ and $\{\alpha^a,\beta_a\}$ their Poincaré dual forms,
\begin{equation}
\int_{\tilde{X}} \alpha^{a}\wedge \beta_b=\delta^a_{~b}\, ,\quad \int_{\tilde{X}}\alpha^{a}\wedge \alpha^b=\int_{\tilde{X}}\beta_{a}\wedge \beta_b=0\, ,\quad a,b=0,...,h^{2,1}(\tilde{X})\, .
\end{equation}
The \textit{periods}
\begin{equation}
z^a=\int_{\Sigma_{(3)a}}\Omega=\int_{\tilde{X}} \Omega\wedge \alpha^a\, ,\quad \mathcal{F}_a=\int_{\Sigma_{(3)}^a}\Omega=\int_{\tilde{X}} \Omega \wedge \beta_a
\end{equation}
form an overcomplete set of coordinates on complex structure moduli space:
locally we have $\mathcal{F}_a=\mathcal{F}_a(z)$ and the $z^a$ are a set of projective coordinates. Similarly, the fluxes $F_3$ and $H_3$ are characterized by the Dirac-quantized flux vectors
\begin{equation}
f^a=\int_{\Sigma_{(3)a}}F_3\, ,\quad f_a=\int_{\Sigma_{(3)}^a}F_3\, ,\quad h^a=\int_{\Sigma_{(3)a}}H_3\, ,\quad h_a=\int_{\Sigma_{(3)}^a}H_3\, ,
\end{equation}
and we will write $\vec{f}=(f_a,f^a)$, $\vec{h}=(h_a,h^a)$.

Prior to orientifolding, the complex structure moduli come in $\mathcal{N}=2$ vector multiplets, and the periods $\mathcal{F}_a(z)$ derive from a prepotential $\mathcal{F}(z)$ via $\mathcal{F}_a(z)=\del_a \mathcal{F}(z)$. The tree-level exact Weil-Petersson metric on complex structure moduli space
is obtained from the K\"ahler potential
\begin{equation}\label{eq:KahlerCS}
K_{cs}=-\ln\left(-i\int_{\tilde{X}} \Omega\wedge \overline{\Omega}\right)=-\ln\bigl(-i\vec{\Pi}^\dagger \Sigma \vec{\Pi}\bigr)\,  ,
\end{equation}
with period vector $\vec{\Pi}=(\del_a \mathcal{F},z^a)^t$ and symplectic pairing $ \Sigma:= \begin{pmatrix}
0 & \id \\
-\id & 0
\end{pmatrix}$.

The orientifold involution induces a splitting of the cohomology groups,
\begin{equation}
H^{p,q}(\tilde{X},\mathbb{Q})=H^{p,q}_+(\tilde{X},\mathbb{Q})\oplus H^{p,q}_-(\tilde{X},\mathbb{Q})
\end{equation}
into even and odd eigenspaces, and the complex structure moduli that survive the projection are counted by
$h^{2,1}_-(\tilde{X},\tilde{\mathcal{I}}):= \text{dim}\, H^{2,1}_-(\tilde{X},\mathbb{Q})$ and come in $\mathcal{N}=1$ chiral multiplets. We will denote these surviving complex structure moduli by $z^a$, $a=1,...,h^{2,1}_-(\tilde{X},\tilde{\mathcal{I}})$.
Likewise, $h^{1,1}_+(\tilde{X},\tilde{\mathcal{I}})$ counts the number of surviving K\"ahler moduli $T^{\alpha}$.  Finally, $h^{1,1}_-(\tilde{X},\tilde{\mathcal{I}})$ and $h^{2,1}_+(\tilde{X},\tilde{\mathcal{I}})$ are the number of axionic chiral multiplets and $U(1)$ vector multiplets, respectively, but will play no role in this paper. The full tree-level effective action has been worked out in \cite{Grimm:2004uq}.

After orientifolding, the K\"ahler potential of eq.~\eqref{eq:KahlerCS} will in general receive corrections from fluxes and orientifold planes, but these are subleading at sufficiently large volume where fluxes are dilute and warping is negligible. The superpotential, however, is exact up to non-perturbative corrections in the K\"ahler moduli and D(-1) instantons and is given by \cite{Gukov:1999ya,Giddings:2001yu}
\begin{align}\label{eq:superpotential}
W(z,\tau,T)&=W_{\text{flux}}(z,\tau)+W_{\text{np}}(z,\tau,T)\,  , \quad \text{with}\\ \sqrt{\tfrac{\pi}{2}}\,W_{\text{flux}}(z,\tau)&=\int_{\tilde{X}} \bigl(F_3-\tau H_3\bigr)\wedge \Omega(z)=\bigl(\vec{f}-\tau \vec{h}\bigr)^t\Sigma \vec{\Pi}(z)\, ,
\end{align}
where $W_{\text{np}}(z,\tau,T)$ parameterizes non-perturbative corrections in the K\"ahler moduli $T$ and the dilaton $\tau$, which are typically difficult to compute.\footnote{The leading order contributions come from D(-1) instantons $\sim e^{2\pi i \tau}$ as well as Euclidean D3-branes and gaugino condensation effects on seven-branes $\sim e^{-2\pi T/c}$, $c\in \mathbb{N}$.} We will neglect these corrections self-consistently. Moreover, even when flux backreaction is severe, the vacuum solutions $D_{\tau}W=D_{z^a}W=0$ obtained using the classical K\"ahler potential \eqref{eq:KahlerCS} are reliable as long as the ten-dimensional geometry is in the supergravity regime, even though the scalar potential away from the supersymmetric minimum can no longer be computed from it \cite{Giddings:2001yu}.

This fact will be particularly important for the purposes of this paper because we will stabilize complex structure moduli near a conifold point in moduli space. In this case, for moderate values of the overall volume modulus, in the vicinity of the (deformed) conifold regions in the Calabi-Yau the fluxes are no longer dilute and their backreaction produces the famous Klebanov-Strassler throats \cite{Klebanov:2000hb}. These produce an exponentially strong gravitational redshift (warping) that is not appropriately captured by eq.~\eqref{eq:KahlerCS} and \eqref{eq:superpotential}. Nevertheless, the Klebanov-Strassler solution falls into the class of imaginary-self-dual (ISD) solutions of \cite{Giddings:2001yu}, and the F-term equations arising from \eqref{eq:KahlerCS} and \eqref{eq:superpotential} can be used as a tool to find points in Calabi-Yau moduli space where the fluxes are indeed ISD.

\subsection{The large complex structure patch}\label{ss:lcs}
In the following, we will be interested in the large complex structure (LCS) patch of complex structure moduli space of $\tilde{X}$, which is mirror dual to the large volume region of the mirror threefold $X$.\footnote{For a recent study of scalar potentials from fluxes in asymptotic limits such as the LCS region, see \cite{Grimm:2019ixq}.}  Let $\{\Sigma_{(2)a}\}$ be a basis of $H_2(X,\mathbb{Z})$ and $\{\Sigma_{(4)}^a\}$ a dual basis of $H_4(X,\mathbb{Z})$, i.e. $\Sigma_{(2)a}\cdot \Sigma_{(4)}^b=\delta_a^{~b}$.
Curve classes $[\mathcal{C}]\in H_2(X,\mathbb{Z})$ are represented by
integer vectors $\beta_a^{\mathcal{C}}$,
\begin{equation}
[\mathcal{C}]=\sum_{a=1}^{h^{1,1}(X)} \beta^\mathcal{C}_a \,[\Sigma_{(2)a}]\,,\qquad \beta^\mathcal{C}_a \in \mathbb{Z}\,.
\end{equation}
The complexified (string frame) curve volumes $z^a=\int_{\Sigma_{(2)a}}\left(B+iJ\right)$, $a=1,...,h^{1,1}(X)$, serve as local coordinates on moduli space and are identified with the type IIB complex structure moduli $z^a$ in a gauge where $z^0=1$. Henceforth, we will work in this gauge and let $a,b=1,...,h^{1,1}(X)$.

The prepotential enjoys the expansion \cite{Hosono:1994av}
\begin{align}\label{eq:Flcs}
\mathcal{F}(z)&=\mathcal{F}_{\text{poly}}(z)+\mathcal{F}_{\text{inst}}(z)\, ,\\
\text{with}\quad \mathcal{F}_{\text{poly}}(z)&=-\frac{1}{3!}\mathcal{K}_{abc}z^az^bz^c+\frac{1}{2}a_{ab}z^az^b+b_az^a+\frac{\chi \zeta(3)}{2(2\pi i)^3}\, .
\end{align}
Here, $\mathcal{K}_{abc}$ are the triple intersection numbers of $X$, and the quadratic term can be taken to be
\begin{equation}
a_{ab}=\frac{1}{2}\begin{cases}
\mathcal{K}_{aab} & a\geq b\\
\mathcal{K}_{abb} & a<b
\end{cases}\, ,
\end{equation}
where $b_a=\frac{1}{24}\int_{\Sigma_{(4)}^a}c_2(X)$, $\chi=\int_{X}c_3(X)$, and $\zeta(3)$ is Ap\'{e}ry's constant. Moreover,
\begin{equation}\label{eq:Finst}
\mathcal{F}_{\text{inst}}(z)=-\frac{1}{(2\pi i)^3} \sum_{[\mathcal{C}]}n_{\mathcal{C}}^0\,\text{Li}_3(q^\mathcal{C})\, ,\quad q^\mathcal{C}:= \exp\left(2\pi i \beta_a^\mathcal{C} z^a\right)\, ,
\end{equation}
where the sum runs over all \emph{effective} curve classes $[\mathcal{C}]$, $n^0_{\mathcal{C}}\in \mathbb{Z}$ are the genus zero Gopakumar-Vafa (GV) invariants \cite{Gopakumar:1998ii,Gopakumar:1998jq}, and $\text{Li}_k(q):= \sum_{n=1}^\infty q^n/n^k$
is the polylogarithm. When all effective curves in $X$ are large, $\mathcal{F}_{\text{inst}}$ parameterizes type IIA worldsheet instanton corrections to the derivatives of the prepotential as
\begin{equation}
\mathcal{F}_a(z)=\del_a\mathcal{F}_{\text{poly}}-\frac{1}{(2\pi i)^2}\sum_{[\mathcal{C}]}n_{\mathcal{C}}^0 \beta_a^{\mathcal{C}}
\,\text{Li}_2(q^\mathcal{C})\,.
\end{equation}

\subsection{Small flux superpotentials at large complex structure}\label{ss:sfs}
As demonstrated in \cite{Demirtas:2019sip}, one can find weakly coupled flux vacua at LCS with exponentially small flux superpotential by making a restricted choice of fluxes.
Near LCS, the flux superpotential splits as
	\begin{equation}\label{eq:superpotentialLCS}
	W_{\text{flux}}(z^a,\tau)\equiv W_{\text{poly}}(z^a,\tau)+W_{\text{inst}}(z^a,\tau)\, ,
	\end{equation}
	where $W_{\text{poly}}(z,\tau)$ is the flux superpotential that arises from the approximation $\mathcal{F}(z)\approx  \mathcal{F}_{\text{poly}}(z)$, and $W_{\text{inst}}(z,\tau)$ parameterizes the instanton corrections,
	\begin{equation}
	-\sqrt{\tfrac{\pi}{2}}W_{\text{inst}}(z,\tau):= (f^a-\tau h^a)\del_a \mathcal{F}_{\text{inst}}(z)+(f^0-\tau h^0)\Bigl(2\mathcal{F}_{\text{inst}}(z)-z^a\del_a\mathcal{F}_{\text{inst}}(z)\Bigr)\, .
	\end{equation}
We choose fluxes
\begin{equation}
\vec{f}=(b_a M^a, a_{ab}M^b,0,M^a)^t\, ,\quad \vec{h}=(0,K_a,0,0)^t\, ,
\end{equation}
parameterized by a pair $\vec{M},\vec{K}\in \mathbb{Z}^{h^{2,1}_-(\tilde{X},\tilde{\mathcal{I}})}$ that satisfies
\begin{equation}
K_a p^a=0\, ,\quad 0\leq -M^aK_a\leq -2Q\, ,\quad \text{with} \quad  p^a:= (\mathcal{K}_{abc}M^c)^{-1}K_b\, ,
\end{equation}
such that $\vec{p}$ is in the K\"ahler cone of $X$. For such a choice of fluxes the polynomial part of the superpotential takes the form
\begin{equation}
W_{\text{poly}}(z,\tau)\propto \frac{1}{2}\mathcal{K}_{abc}M^az^bz^c-\tau K_az^a\, ,
\end{equation}
and has a supersymmetric valley $\del_{z^a}W_{\text{poly}}=W_{\text{poly}}=0$ along the one-dimensional locus where $z^a=p^a\tau$. Generically, the orthogonal directions to the flat valley are heavy and can be integrated out --- we will verify this in explicit examples in \S\ref{sec:explicit}.

As the polynomial part of the superpotential vanishes along the flat valley, the instanton corrections to the superpotential become relevant, and serve to stabilize $\tau$. The effective superpotential is
\begin{align}
W_{\text{eff}}(\tau):=W_{\text{inst}}(p^a\tau,\tau)=&\sqrt{\tfrac{2}{\pi}}\frac{1}{(2\pi i)^2}\sum_{[\mathcal{C}]}n_{\mathcal{C}}^0 M^a\beta_a^{\mathcal{C}} \,\text{Li}_2\left(e^{2\pi i \beta_a^{\mathcal{C}}p^a\tau}\right)+\mathcal{O}(e^{2\pi i \tau},e^{-2\pi T})\, .
\end{align}
In the regime where D(-1) instanton effects can be consistently neglected, one should only retain terms in the sum with $\beta_a^{\mathcal{C}}p^a<1$. For appropriately aligned $p^a$ and suitably hierarchical GV invariants the above structure leads to a \textit{racetrack} stabilization of $\tau$ at weak coupling and near LCS.

\section{Stabilizing near the conifold}\label{sec:result}
We will now extend the construction of \cite{Demirtas:2019sip} to operate in a regime where one or more of the moduli are away from their LCS region, and are instead exponentially close to developing a conifold singularity.

At a conifold singularity in complex structure moduli space a collection of $n_{\text{cf}}$ three-cycles shrink to zero size \cite{Candelas:1989js,Candelas:1989ug}. Let us assume that these all lie in the same homology class, corresponding to one of the basis elements $\Sigma_{(3)a}$, with corresponding modulus $\zc$.  We denote the remaining moduli by $z_i$, i.e.~$\{z^a\}=\{\zc, z^i\}$.
Then the dual period, which we will denote by $\mathcal{F}_{\text{cf}}$, takes the form
\begin{equation}\label{eq:Bcycle_period}
\mathcal{F}_{\text{cf}}(z^a)=\mathcal{F}_{\text{cf}}(z^i, \zc)= \frac{n_{\text{cf}}}{2\pi i}\zc\ln(\zc)+f(z^i,\zc)\, ,
\end{equation}
where $f(z^i,\zc)$ is a model-dependent function. Although $f(z^i,\zc)$ is holomorphic around $\zc=0$, it will play an important role in our discussion, because generically $f(z^i,0) \neq 0$.

Then, with $-M$ units of $F_3$ flux on the shrinking cycle and $K$ units of $H_3$ flux on the dual cycle, the flux superpotential splits as \cite{Giddings:2001yu}
\begin{align}\label{eq:conifold_superpotential}
W_{\text{flux}}&= W_{\text{cf}}(z^a)+W_{\text{bulk}}(z^a)\\
\sqrt{\tfrac{\pi}{2}}W_{\text{cf}}(z^a)&:= M\left(\frac{n_{\text{cf}}}{2\pi i}\zc\ln(\zc)+f(z^a)\right)-\tau K \zc\, ,
\end{align}
where $W_{\text{bulk}}(z^a)$ is holomorphic around $\zc=0$ and parameterizes the contribution to the superpotential from other cycles.
Provided that $|K|>g_s |M|$ and that $K$ and $M$ have the same sign, this stabilizes the conifold modulus exponentially close to the singularity,
\begin{equation}
|\zc|\sim \mathrm{exp}\Biggl(-\frac{2\pi K}{n_{\text{cf}}g_sM}\Biggr)\, .
\end{equation}
Upon stabilizing in this regime, one is left with codimension-three defects hosting confining Klebanov-Strassler gauge theories if $g_sM \ll 1$, or with warped throats with a controlled ten-dimensional supergravity description in the regime $g_sM \gg 1$ \cite{Klebanov:2000hb}. In the former case, $|\zc|^{1/3}$ is identified with the confining scale of the gauge theory, while in the latter case it is the gravitationally redshifted Randall-Sundrum-type \cite{Randall:1999ee} warp factor\footnote{Here we assume only a moderately large Calabi-Yau volume $\mathcal{V}$, i.e. $|\zc|\ll 1$ and also $\mathcal{V}|\zc|^2\ll 1$. In the opposite regime of parametrically large volume such that $\mathcal{V}|\zc|^2\gg 1$, there is neither a throat nor a gauge theory but simply an everywhere weakly curved conifold region with dilute fluxes. For results on moduli stabilization in this opposing regime, see \cite{Blumenhagen:2016bfp}.} $e^{A}|_{\mathrm{tip}}\sim |\zc|^{1/3}$.

For generating uplifts to de Sitter vacua by including anti-D3-branes it is natural to consider the regime $g_sM \gtrsim 1$ where the infrared region of the throat supports metastable supersymmetry-breaking anti-D3-brane configurations that contribute to the vacuum energy \cite{Kachru:2003aw}. However, this restriction might well be unnecessary due to the plausible existence of a supersymmetry-breaking vacuum in the Klebanov-Strassler gauge theory, see e.g. \cite{Argurio:2006ny,Retolaza:2015nvh,Argurio:2019eqb,Argurio:2020dkg,Argurio:2020npm}.
In any event, in this paper we will study classical conifold vacua in both regimes of $g_s M$, and defer metastable supersymmetry breaking to future work.

After stabilizing exponentially close to the conifold singularity, the flux superpotential reads
\begin{equation}
W_{\text{flux}}(z^i,\zc)= \sqrt{\tfrac{2}{\pi}}Mf(z^i,0)+W_{\text{bulk}}(z^i,0)  +\mathcal{O}(\zc)\, .
\end{equation}
In particular, the holomorphic piece $f(z^a)$ in the conifold period \eqref{eq:Bcycle_period} gives an $\mathcal{O}(1)$ contribution to the superpotential that has to be canceled against the bulk superpotential $W_{\text{bulk}}$ to give a small flux superpotential, as alluded to in the introduction. We will now explain how this cancellation is achieved.

\subsection{The conifold prepotential from a shrinking curve}

In this section we will compute the periods of $\tilde{X}$ analytically
near certain types of conifold points.  The idea is to analytically continue the periods computed at LCS into the regime where one of the moduli, $\zc$, is small and close to a conifold singularity, while the other moduli $z^i$ stay large.
Our result will take the form of a double expansion in the conifold modulus $\zc \ll 1$ and in type IIA worldsheet instantons wrapping curves in $X$ that are not mirror-dual to the A-cycle of the conifold.

Performing an analytic continuation from the LCS region to the conifold region generally requires knowing the instanton expansion of the prepotential in eq.~\eqref{eq:Finst} to all orders.  After all, the conifold branch cut can at best arise at the radius of convergence of the LCS expansion.
Lacking such all-orders information, one might instead determine the prepotential to high order in the instanton expansion, and compute the Taylor coefficients of $f(z^a)$ in \eqref{eq:Bcycle_period} numerically by comparing in an overlapping region where both expansions converge, as in \cite{Blumenhagen:2016bfp}. However, in order to demonstrate an accurate cancellation between the bulk and conifold superpotential \eqref{eq:fine-tuning} by this approach, one has to reach high numerical precision.  For this reason we opt instead for an analytic approach, and now lay out a set of conditions under which we can obtain the required all-orders information.

\begin{figure}
	\centering
	\begin{tabular}{p{0.7cm} p{7cm}}
		\begin{center}
			
			\begin{tabular}{l}
				$n^0_{(i,5,0)}$ \vspace{0.2cm}\\
				$n^0_{(i,4,0)}$ \vspace{0.2cm}\\
				$n^0_{(i,3,0)}$ \vspace{0.2cm}\\
				$n^0_{(i,2,0)}$ \vspace{0.2cm}\\
				$n^0_{(i,1,0)}$ \vspace{0.2cm}\\
				$n^0_{(i,0,0)}$ \vspace{0.2cm}
			\end{tabular}
			
		\end{center}
		&
		\begin{center}
			\vspace{0.2cm}
			\includegraphics[keepaspectratio,height=4.0cm,origin=c]{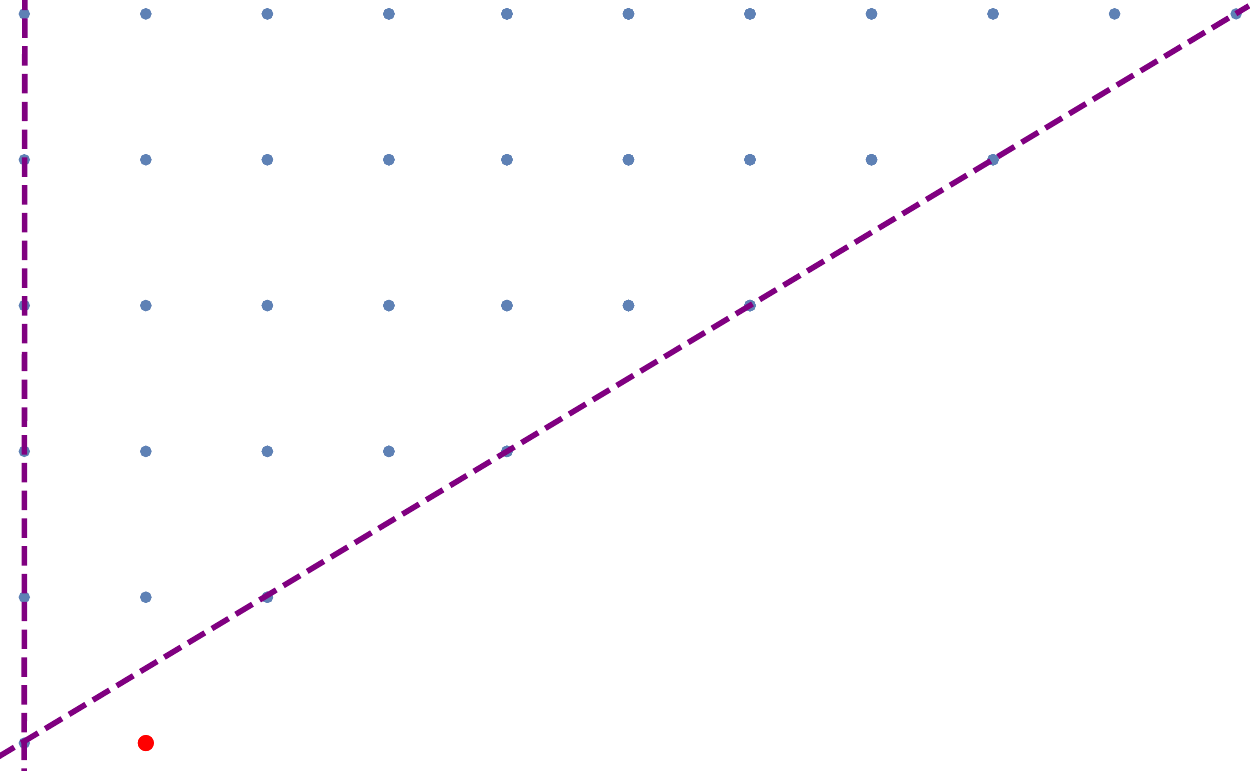}
		\end{center}	
	\end{tabular}
	$\overset{z^1\rightarrow 0}{\longrightarrow}$
	\begin{tabular}{p{1.5cm} p{1cm}}
		\begin{center}
			
			\begin{tabular}{r}
				$\sum_{i=0}^{10}n^0_{(i,5,0)}$ \vspace{0.2cm}\\
				$\sum_{i=0}^{8}n^0_{(i,4,0)}$ \vspace{0.2cm}\\
				$\sum_{i=0}^{6}n^0_{(i,3,0)}$ \vspace{0.2cm}\\
				$\sum_{i=0}^{4}n^0_{(i,2,0)}$ \vspace{0.2cm}\\
				$\sum_{i=0}^{2}n^0_{(i,1,0)}$ \vspace{0.2cm}\\
				$n^0_{(1,0,0)}$ \vspace{0.2cm}
			\end{tabular}
			
		\end{center}
		&
		\begin{center}
			\vspace{0.2cm}
			\includegraphics[keepaspectratio,height=4.1cm,origin=c,clip,trim= 6cm 0cm 6cm 0.2cm]{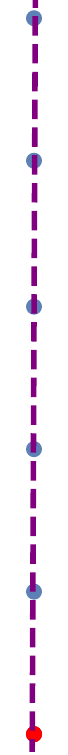}
		\end{center}	
	\end{tabular}
	\caption{The slice $\beta^{\mathcal{C}}_3=0$ in the Mori cone of the Calabi-Yau $X$ described in \S\ref{sec:example}. Blue lattice points are populated by nonvanishing GV invariants $n^0_{(i,j,0)}$.   The red lattice point is GV-nilpotent of order one, and lies outside the closure of the interior cone (bounded by dashed purple lines). Near the origin of the Coulomb branch $z^1\rightarrow 0$, one retains a controlled expansion in $e^{2\pi i z^{2,3}}$ with coefficients $\sum_i n^0_{(i,j,k)}$. These are the GV invariants on the Higgs branch \cite{Candelas:1993dm}, and are computable because the sum over $i$ terminates: each row in the figure has finite length.}
	\label{fig:GVlattice_collapsed}
\end{figure}

First, let us introduce some terminology. We write $\mathcal{M}(X)$ for the Mori cone of $X$, and we call a curve class $[\mathcal{C}] \in \mathcal{M}(X)$ \textit{GV-nilpotent of order $k_0$} if the genus zero GV invariants of $k[\mathcal{C}]$ vanish for all $k>k_0$. A class that is not GV-nilpotent of order $k_0$ for any finite $k_0$ we call \textit{GV-potent}.  We define the \textit{interior cone} as the closure of the real cone generated by all GV-potent curve classes in $\mathcal{M}(X).$

Comparing to eq.~\eqref{eq:Finst}, one sees that the infinite tower of instanton corrections from worldsheet instantons wrapping a GV-nilpotent curve class $[\mathcal{C}]$ an arbitrary number of times are simply determined by a \textit{finite} sum over polylogarithms. Therefore, if we can find a slice in moduli space where \textit{only} GV-nilpotent curve classes shrink, we only need to analytically continue the well-known polylogarithms. The following condition (which we will establish in the example of \S\ref{sec:example}) guarantees this.

We suppose that there exists a curve $\mathcal{C}_{v}\in \mathcal{M}(X)$ in the mirror threefold $X$ that is GV-nilpotent of order one and lies outside the closure of the cone generated by all other curve classes
in $\mathcal{M}(X)$ with non-vanishing GV invariants: see Figure \ref{fig:GVlattice_collapsed}. (In particular, $\mathcal{C}_{v}$ lies outside the interior cone.)
In this case, there exists a wall $\mathcal{W}_{\mathcal{C}_v}$ of the K\"ahler cone of $X$ where the volume of $\mathcal{C}_v$ vanishes. The wall $\mathcal{W}_{\mathcal{C}_v}$ is a cone itself, and asymptotically far out in this cone the volumes of all other curves with non-vanishing GV invariants tend to infinity. For ease of exposition we will assume that $\mathcal{C}_v$ is an element of our basis of curves, so without loss of generality we can choose $\mathcal{C}_{v}=\Sigma_{(2)1}$. We will denote the corresponding K\"ahler modulus of $X$ by $\zc:= z^1$. The other K\"ahler moduli of $X$ will be denoted by $z^i$ with $i=2,...,h^{1,1}(X)$. Then, we have
\begin{align}
\mathcal{F}_{i}(\zc,z^{i}):=\del_{z^i}\mathcal{F}(\zc,z^{i})=&-\frac{1}{2}\mathcal{K}_{iab}z^a z^b+a_{ia}z^{a}+b_i-\frac{1}{(2\pi i)^2}\sum_{[\mathcal{C}]\neq [\mathcal{C}_v]}n_{\mathcal{C}}^0 \beta_i^{\mathcal{C}}\,\text{Li}_2(q^\mathcal{C})\, .
\end{align}
Since the vanishing class $[\mathcal{C}_v]$ does not appear in the instanton sum, the arguments of the polylogarithm remain small as in the LCS regime even for small $\zc$. This is not quite enough for a sensible expansion:  for two curve classes $[\mathcal{C}]$ and $[\mathcal{C}']$ with $[\mathcal{C}]-[\mathcal{C}']\propto [\mathcal{C}_v]$ the corresponding arguments of the polylogarithms become identical in the limit $\zc\rightarrow 0$, so the GV invariants of curve classes differing by the vanishing class are effectively summed up in that limit.\footnote{Note that if there is a Higgs branch meeting the origin of the Coulomb branch at $\zc=0$, such a summation yields the GV invariants of the Higgs branch \cite{Candelas:1993dm}.}
For the instanton expansion to remain controlled, we need that these sums of GV invariants terminate. But this is guaranteed because our vanishing curve class lies \textit{outside} the interior cone (see Figure \ref{fig:GVlattice_collapsed}). Thus, any ray parallel to $[\mathcal{C}_v]$ intersects finitely many lattice points in the interior cone.

The remaining task is to compute $\mathcal{F}_{1}(\zc,z^{i}):=\del_{\zc}\mathcal{F}(\zc,z^i)$. The crucial difference compared to the $\mathcal{F}_{i}(\zc,z^{i})$ is that the vanishing curve \textit{does} contribute to the instanton sum, so we need to evaluate the polylogarithm near $q=1$. This is most easily done using Euler's reflection formula\footnote{The corresponding series of instanton corrections to the period vector is thus resummed into the perturbative one-loop correction from integrating out light hypermultiplets from wrapped D2-branes/M2-branes near the origin of the Coulomb branch \cite{Strominger:1995cz,Greene:1995hu,Gopakumar:1998ii,Gopakumar:1998jq}.}
\begin{equation}\label{eq:reflection_formula}
-\frac{\text{Li}_2(e^{2\pi i z})}{(2\pi i )^2}=\frac{1}{24}+\frac{z}{2\pi i}\ln\bigl(1-e^{2\pi i z}\bigr)+\frac{\text{Li}_2(1-e^{2\pi i z})}{(2\pi i )^2}=\frac{1}{24}+\frac{z}{2\pi i}\Bigl(\ln(-2\pi i z)-1\Bigr)+\mathcal{O}(z^2)\, .
\end{equation}
Thus, finally, we arrive at
\begin{align}\label{eq:Bcycle_period_explicit}
\mathcal{F}_{1}(\zc,z^{i}):=\del_{\zc}\mathcal{F}(\zc,z^i)=&~n_{\text{cf}}\frac{\zc}{2\pi i}\ln(1-e^{2\pi i \zc})\\
-\frac{1}{2}\mathcal{K}_{1ab}z^a z^b+a_{1a}z^{a}+b_1+n_{\text{cf}}&\left(\frac{1}{24}+\frac{\text{Li}_2(1-e^{2\pi i \zc})}{(2\pi i)^2}\right)-\frac{1}{(2\pi i)^2}\sum_{[\mathcal{C}]\neq [\mathcal{C}_v]}n_{\mathcal{C}}^0 \beta_1^{\mathcal{C}}\,\text{Li}_2(q^\mathcal{C})\, .\nonumber
\end{align}
Provided that the singular locus $\zc\rightarrow 0$ is in fact a conifold singularity in $\tilde{X}$, we may identify $\mathcal{F}_{\text{cf}}(\zc,z^i)\equiv\mathcal{F}_1(\zc,z^i)$.
The first term in \eqref{eq:Bcycle_period_explicit} contains the universal logarithm of the general conifold period of eq.~\eqref{eq:Bcycle_period}, as well as an infinite series of holomorphic corrections. These corrections, together with the entire second line of eq.~\eqref{eq:Bcycle_period_explicit}, constitute the holomorphic term $f(z^a)$ in \eqref{eq:Bcycle_period}.  We can therefore compute $f(z^a)$ to any desired accuracy by computing the GV invariants $n_{\mathcal{C}}^0$ of curve classes $[\mathcal{C}]$ up to a sufficently high degree.

For completeness, we note that one similarly obtains a systematic expansion of the prepotential around the conifold locus, from which the period $\mathcal{F}_0$ can be computed via $\mathcal{F}_0=2\mathcal{F}-z^a \mathcal{F}_a$. First, one uses a version of Euler's reflection formula for $\text{Li}_3(e^{2\pi i z})$,
\begin{equation}\label{eq:reflection_formula_3}
-\frac{\text{Li}_3(e^{2\pi i z})}{(2\pi i)^3}=\frac{z^2}{4\pi i} \ln(-2\pi i z)-\frac{1}{(2\pi i)^3}\sum_{n=0}^\infty \frac{\hat{\zeta}(3-n)}{n!}(2\pi iz)^n\, ,
\end{equation}
with $\hat{\zeta}(k):=\zeta(k)$ for $k\neq 1$ and $\hat{\zeta}(1):=\frac{3}{2}$. The relation \eqref{eq:reflection_formula_3} follows directly from an identity obeyed by the Lerch transcendent $\Phi(e^{2\pi i z}, 3, 1)$ \cite{Bateman:100233,GradshteynRyzhik}.
Using this, we may expand the prepotential systematically around $\zc=0$,
\begin{equation}
\mathcal{F}(\zc,z^i)=n_{\text{cf}}\frac{\zc^2}{4\pi i}\ln(-2\pi i \zc)+\sum_{n=0}^\infty \frac{\mathcal{F}^{(n)}(z^i)}{n!}\, \zc^n \, ,
\end{equation}
with
\begin{equation}
	\mathcal{F}^{(n)}(z^i):=P^{(n)}(z^i)- \frac{n_{\text{cf}}}{(2\pi i)^{3-n}}\, \hat{\zeta}(3-n)-\frac{1}{(2\pi i)^{3-n}}\sum_{[\mathcal{C}]\neq [\mathcal{C}_v]}n_{\mathcal{C}}^0\left(\beta_1^{\mathcal{C}}\right)^n\, \, \left.\text{Li}_{3-n}(q^\mathcal{C})\right|_{\zc=0}\, .
\end{equation}
Here, the $P^{(n)}(z^i):=\left.\left(\del_{\zc}^n \mathcal{F}_{\text{poly}}\right)\right|_{\zc=0}$ are degree $3-n$ polynomials in the bulk moduli $z^i$.

In summary, we have gained the much needed analytical control over the period vector near a special class of conifolds. For this special class the shrinking $S^3$ of the conifold is mirror dual to a shrinking \textit{curve}. This is a nontrivial restriction because a conifold singularity in $\tilde{X}$ may be mirror dual to a shrinking curve, divisor or entire threefold $X$.\footnote{For instance the famous conifold point in the mirror quintic is mirror dual to a shrinking Calabi-Yau threefold \cite{Candelas:1990rm}. For more examples, see \cite{Grimm:2008ed,Blumenhagen:2008zz}.}

\subsection{Moduli stabilization in three steps}

We will turn on the following subset of three-form fluxes:
\begin{equation}\label{eq:restricted_fluxes}
\vec{f}=(P_0,P_a,0,M^a)^t\, ,\quad \vec{h}=(0,K_a,0,0)^t\, .
\end{equation}
Instead of splitting the superpotential into contributions from the conifold fluxes and the bulk, we expand in the conifold modulus $\zc$,
\begin{equation}\label{eq:superpotential_split}
\sqrt{\tfrac{\pi}{2}}\,W_{\text{flux}}(\zc,z^i,\tau)=W^{(0)}(z^i,\tau)+W^{(1)}(z^i,\tau,\zc)\zc +\mathcal{O}(\zc^2)\, .
\end{equation}
The $\mathcal{O}(\zc^0)$ term takes a form akin to the LCS expansion \eqref{eq:superpotentialLCS}, but only in the bulk moduli $z^i$, i.e.
\begin{equation}\label{eq:superpotential_conifold_expansion}
W^{(0)}(z^i,\tau)=W^{(0)}_{\text{poly}}(z^i,\tau)+W^{(0)}_{\text{inst}}(z^i)\,,
\end{equation}
with
\begin{align}
&W^{(0)}_{\text{poly}}(z^i,\tau)=M^a\left(\frac{1}{2}K_{aij}z^i z^j-a_{ai}z^i-\tilde{b}_a\right)+P_i z^i+P_0-\tau K_iz^i\, ,\\
&W^{(0)}_{\text{inst}}(z^i)=\frac{1}{(2\pi i)^2}\sum_{[\mathcal{C}]\neq [\mathcal{C}_v]}n^0_{\mathcal{C}} M^a \beta^{\mathcal{C}}_a\, \text{Li}_2(q^\mathcal{C})|_{\zc=0}\, ,
\end{align}
and with shifted $\tilde{b}_a:= b_a+n_{\text{cf}}\delta_{a1}/24$. Note that by expanding in $\zc$ we have absorbed all contributions to the flux superpotential that survive in the conifold limit $\zc \to 0$ into the expression $W^{(0)}(z^i,\tau)$ containing both bulk and conifold contributions.
Provided that we can stabilize the bulk moduli and the dilaton such that
$\vert\langle W^{(0)}(z^i,\tau) \rangle\vert \ll 1$,
and also stabilize $\zc$ near the conifold, a small \textit{overall} flux superpotential as in eq.~\eqref{eq:fine-tuning} will result.

The next-to-leading order term in the expansion \eqref{eq:superpotential_conifold_expansion} contains the conifold logarithm
\begin{equation}
W^{(1)}(z^i,\tau)=M \frac{n_{\text{cf}}}{2\pi i}\Bigl(\ln(-2\pi i \zc)-1\Bigr)-\tau K
+\mathcal{K}_{1ai}M^az^i+P-a_{1a}M^a +\mathcal{O}\left(e^{2\pi i z^i}\right) \, ,
\end{equation}
where
\begin{equation}
M:= -M^1\, ,\quad  P:= P_1 \, ,\quad \text{and} \quad K:= K_1\, .
\end{equation}
Given this structure, we may stabilize moduli in three essentially independent steps. First, we use $W^{(0)}_{\text{poly}}$ to stabilize all but one combination of the bulk moduli and dilaton as in \cite{Demirtas:2019sip} in a vacuum with $W^{(0)}_{\text{poly}}=0$. For that we choose integer fluxes such that
\begin{equation}\label{eq:quantization_condition}
P_{i}=a_{i a}M^a\, ,\quad  P_0= \tilde{b}_a M^a\, ,
\end{equation}
and
\begin{equation}\label{eq:flat_Kfluxes}
K_i p^i=0\, ,\quad 0\leq -M^aK_a\leq 2Q\, ,\quad \text{with} \quad  p^i:= (\mathcal{K}_{ija}M^a)^{-1}K_j\, ,
\end{equation}
such that $p^i$ is interior to the wall of the K\"ahler cone of $X$ specified by $\zc=0$. In analogy to the previous section, this stabilizes the \textit{bulk} moduli along the valley $z^i=p^i\tau$ and indeed $W^{(0)}_{\text{poly}}(z^i=p^i\tau,\tau)=0$. At this stage the leading contributions to the superpotential from the bulk and the conifold have been canceled against each other perfectly. The neglected contributions from $W^{(0)}_{\text{inst}}$ and $\zc W^{(1)}$ to the F-term conditions of the directions orthogonal to the flat valley $z^i=p^i\tau$ are negligible if $\text{Im}(\tau)$ is large and the conifold modulus $\zc$ is small.

We note that the above corresponds to a Wilsonian integrating out of heavy degrees of freedom (the orthogonal moduli), so we can stabilize the light degrees of freedom $(\tau,\zc)$ using the low energy effective field theory. Below the mass scale of the heavy orthogonal moduli, the effective superpotential reads
\begin{equation}
\sqrt{\tfrac{\pi}{2}}W_{\text{eff}}(\tau,\zc)=W^{(0)}_{\text{inst}}(z^i=p^i\tau,\tau)+\zc W^{(1)}(z^i=p^i\tau,\tau,\zc)\, .
\end{equation}
Next, we may solve the F-term equation of the conifold modulus $\zc$, giving
\begin{equation}\label{eq:conifold_vev}
|\zc|=\frac{1}{2\pi}\,\mathrm{exp}\Biggl(-\frac{2\pi K'}{n_{\text{cf}}g_s M}\Biggr)+\mathcal{O}\left(\zc^2,\zc e^{2\pi i p^i\tau}\right)\, ,
\end{equation}
in terms of the string coupling $1/g_s=\text{Im}(\tau)$, and with $K':= K_1-M^a\mathcal{K}_{1ai}p^i$.\footnote{The combination $K'/n_{\text{cf}}$ appearing in eq.~\eqref{eq:conifold_vev} is naturally interpreted as the integrated (but not necessarily quantized) three-form field strength residing in a single throat.  The presence of more than one throat that share the same B-cycle may lead to the presence of light degrees of freedom (\textit{thraxions}) that control the relative distribution of fluxes \cite{Hebecker:2018yxs} and would threaten the stability of a warped uplift. In our example of \S\ref{sec:example} we will have $n_{\text{cf}}=2$, but the two throats are identified in the orientifold so the thraxion is projected out.} The phase of $\zc$ is similarly stabilized in terms of $C_0=\text{Re}(\tau)$.
As long as $\zc\ll 1$ the stabilization of $\zc$ does not affect the stabilization of the previously integrated-out heavy moduli.

The remaining light direction $\tau$ can be stabilized as in \cite{Demirtas:2019sip} using the instanton corrections $W^{(0)}_{\text{inst}}$ in a regime where $\text{Im}(\tau)$ is indeed large.\footnote{We are primarily interested in the regime $|\zc|^{2/3}\sim |W_0|\ll 1$ where $\tau$ is much lighter than the conifold modulus, i.e. $m_{\tau}\sim |W_0|\sim |\zc|^{2/3}\ll |\zc|^{1/3}\sim m_{\zc}$. In the opposite regime $|\zc|^{1/3}\ll |W_0|$ it is more natural in the Wilsonian sense to first integrate out $\tau$ and finally stabilize the light conifold modulus $\zc$, but our formulas are valid in both regimes.}
If we stabilize in a regime where the resulting vev $\langle W^{(0)}_{\text{inst}}\rangle$ is much larger than $\zc W^{(1)}$ then it is consistent to neglect the contribution of $\zc  W^{(1)}$ to the F-term of $\tau$.
The vacuum value of the full flux superpotential is then given, up to corrections of $\mathcal{O}(\zc)$, by
\begin{equation}
\boxed{\Biggl.\Biggr.W_0 \approx \sqrt{\tfrac{2}{\pi}}\Bigl\langle W^{(0)}_{\text{inst}}(p^i\tau,\tau)\Bigr\rangle \, .}
\end{equation}
We have therefore extended the mechanism of \cite{Demirtas:2019sip} to conifold vacua.

\subsection{Comments on the supergravity approximation}
In the regime of small string coupling $g_s<1$, the curvature of the infrared region of the throat is large in string units unless one chooses sufficiently large $M$ to ensure that $g_sM>1$. Furthermore, obtaining a substantial throat hierarchy $|\zc|\ll 1$ requires choosing $K>n_{\text{cf}}\,g_s M>1$, and thus the contribution to the D3-brane tadpole from the throats is generically substantial, $KM\gg 1$. This leaves little room to choose appropriate bulk fluxes that would give rise to a small flux superpotential. Given a flux tadpole $Q$ the maximum possible value of $g_sM$ is obtained by saturating the above inequalities, which gives
\begin{equation}\label{eq:KSradius_upper_bound}
g_sM|_{\text{max}}\lesssim \sqrt{|Q|}\, .
\end{equation}
In \S\ref{sec:example} we will work with an orientifold that has $Q=-52$, so that the maximum possible $g_sM$ is of order $7$.  In this orientifold we will be able to find fluxes giving $g_sM\sim 3$.
Whether this is enough for the anti-D3-brane stability analysis of \cite{Kachru:2002gs} to apply will be left for future work to decide.

We note that it has been argued in \cite{Bena:2018fqc} that the restriction \eqref{eq:KSradius_upper_bound} generally prevents one from obtaining well-controlled warped throats in weakly-coupled type IIB compactifications.
While it is clear that parametrically large $g_sM$ is impossible, we see no reason why finding numerically large values should be impossible.

\section{An example}\label{sec:example}

In this section we will construct explicit flux vacua with small flux superpotential, exponentially close to a conifold singularity, along the lines discussed in the previous section.

We start with a certain Calabi-Yau hypersurface $X$ in a toric fourfold $Y$, with Hodge numbers $h^{1,1}(X)=3$ and $h^{2,1}(X)=99$, and specify an O3/O7 orientifold involution $\mathcal{I}:X\rightarrow X$ such that $h^{1,1}_-(X,\mathcal{I})=h^{2,1}_+(X,\mathcal{I})=0$. Using the Greene-Plesser description \cite{Greene:1990ud} of the mirror threefold $\tilde{X}$ as the resolution of an orbifold $X/G$ for a particular abelian group $G$, we show that the induced action of $\mathcal{I}$ on $\tilde{X}$, denoted $\tilde{\mathcal{I}}:\tilde{X}\rightarrow \tilde{X}$, specifies an O3/O7 orientifold involution in $\tilde{X}$ with $h^{1,1}_-(\tilde{X},\tilde{\mathcal{I}})=h^{2,1}_+(\tilde{X},\tilde{\mathcal{I}})=0$.
For the orientifold of $\tilde{X}$ specified by $\tilde{\mathcal{I}}$ we will find conifold vacua with small flux superpotential.\footnote{Alternatively, one can search for vacua along the $G$-symmetric locus in the complex structure moduli space of the orientifold of $X$, as in \cite{Giryavets:2003vd}. This would require an analysis of the action of $G$ on the three-cycles in $X$, which we would like to avoid in this paper.}

For simplicity, we choose to cancel the D7-brane tadpole locally by placing four D7-branes on top of each O7-plane, with trivial gauge bundle. More precisely, we choose diagonal worldvolume fluxes $F_{D_I}=-\frac{1}{2}c_1(D_I)$ on the worldvolumes of the $\mathfrak{so}(8)$ seven-brane stacks on the fixed divisors $D_I$ to cancel potential Freed-Witten anomalies \cite{Freed:1999vc}. Furthermore, we turn on a half-integral orientifold-even NSNS two-form,
\begin{equation}
B_2=\sum_I \frac{1}{2}[D_I]\, .
\end{equation}
Since $c_1(D_I)= -i^*[D_I]$, where $i^*$ is the pull-back of the two-form $[D_I]$ to the divisor $D_I$, the gauge-invariant field-strengths $\mathcal{F}_{D_I}:= F_{D_I}-i^*B_2$ are proportional to the Poincar\'{e} duals of the two-cycles in $D_I$ obtained by intersecting with $\sum_{J\neq I}[D_I]$. Because the orientifold-invariant Calabi-Yau will be smooth, the O7-planes do not intersect each other. Therefore, the field-strengths $\mathcal{F}_{D_I}$ are trivial in $H^2(D_I,\mathbb{Z})$ and so do not contribute to the D3-brane and D5-brane tadpoles. In this configuration the total D3-brane charge $Q$ of the seven-brane stacks and O3-planes is given by
\begin{equation}
Q=-\frac{\chi(\mathfrak{F}_{\mathcal{I}})}{4}\, ,
\end{equation}
where $\chi(\mathfrak{F}_\mathcal{I})$ is the Euler characteristic of the fixed locus $\mathfrak{F}_\mathcal{I}$ of $\mathcal{I}$ \cite{Collinucci:2008pf}.
\subsection{A Calabi-Yau threefold and an orientifold}\label{ss:x}
Let $\Delta^\circ\subset N\simeq \mathbb{Z}^4$ be the favorable reflexive polytope whose points not interior to facets are the columns of
\begin{equation}\label{eqn:N_points}
\begin{pmatrix}
-1 & 1 & -1  & -1 & -1  & -1 & -1 \\
3 & -1 & 0   & 0  & 0   & 0  & 0  \\
-2 & 0 & 0   & 0  & 1   & 2  & 1  \\
-1 & 0 & 1   & 0  & 1   & 0  & 0
\end{pmatrix}\, .
\end{equation}
The first six of these points are vertices. A fine, regular, star triangulation (FRST) of the points in \eqref{eqn:N_points} defines a complete, simplicial fan. The toric fourfold $Y$ defined by this fan contains a smooth anticanonical hypersurface $X$ that is Calabi-Yau. The linear relations among these points define the rows of a gauged linear sigma model (GLSM) charge matrix
\begin{equation}\label{eqn:GLSM}
Q=\begin{pmatrix}
0 & 0 & 1 & -1 & -1 & 0 & 1 \\
1 & 3 & -1 & 1 & 2 & 0 & 0 \\
0 & 0 & 0 & 1 & 0 & 1 & -2
\end{pmatrix}\, .
\end{equation}
Each of the points in \eqref{eqn:N_points} corresponds to a prime effective divisor $\widehat{D}_i \subset Y$ defined by $x_i=0$, that intersects $X$ transversely. The polytope $\Delta^\circ$ has three FRSTs, each giving rise to a smooth Calabi-Yau threefold with favorable embedding in $Y$, i.e.~the divisors $D_i := \widehat{D}_i \cap X$ generate $H_4(X,\mathbb{Z})$\cite{Batyrev:1994hm}.
Denoting the FI parameters by $\xi^a$, we find that the corresponding K\"ahler cones are\footnote{We have computed the K\"ahler cone and intersection numbers using the software package {\tt{CYTools}} \cite{CYtools}.}
\begin{align}
\text{CY}_1:& \quad \,\,\phantom{-} \xi^1>0\, ,\quad \,\xi^2>0\, ,\quad \quad \quad \quad \,\, \xi^3>0\, ,\\
\text{CY}_2:& \quad -\xi^1>0\, , \quad \xi^2+2\xi^1>0\, ,\quad \phantom{-}\xi^3+\xi^1>0\, ,\\
\text{CY}_3:& \quad \phantom{-}\,\,\,\xi^3>0\, , \quad \xi^2+2\xi^1>0\, ,\quad -\xi^3-\xi^1>0\, .
\end{align}
One can show that the triple intersection numbers and the second Chern class of the three phases agree with each other in a suitable basis. Thus, it follows from Wall's theorem \cite{WALL1966} that the three Calabi-Yau manifolds are all diffeomorphic to each other. Without loss of generality we will focus on the phase $\text{CY}_1$.

In general, computing the K\"ahler cone (or its dual, the Mori cone) of a Calabi Yau hypersurface is difficult. However, in the present example the Mori cone of the ambient fourfold $Y$, $\mathcal{M}(Y)$, is equivalent to the Mori cone of the Calabi-Yau hypersurface $\mathcal{M}(X)$. This can be shown as follows. We have $\mathcal{M}(Y) \supseteq \mathcal{M}(X)$ as $X$ is a holomorphic submanifold of $Y$. The GV invariants of the generators of $\mathcal{M}(Y)$ are non-trivial (see \eqref{eq:GVinvariants}), so $\mathcal{M}(X) \supseteq \mathcal{M}(Y)$.

The Stanley-Reisner (SR) ideal of this phase is\footnote{We have used {\tt{Sage}} to determine this \cite{sagemath}.}
\begin{equation}
SR=\{x_3x_6,\,x_4x_6,\,x_3x_7,\,x_1x_2x_4x_5,\,x_1x_2x_5x_7\}\, .
\end{equation}
We choose to work in a basis of divisor classes $H^a\in H_2(X,\mathbb{Z})$ dual to the generators $\mathcal{C}_a\equiv [\Sigma_{2,a}]$ of $\mathcal{M}(X)$, i.e.
\begin{equation}
H^1=[D_1]+[D_3]\, ,\quad H^2=[D_1]\, ,\quad H^3=[D_6]\, .\label{eqn:mori basis}
\end{equation}
In this basis, the non-vanishing triple intersection numbers are
\begin{equation}
\mathcal{K}_{111}=\mathcal{K}_{112}=\mathcal{K}_{122}=4\, ,\quad \mathcal{K}_{113}=\mathcal{K}_{123}=\mathcal{K}_{223}=2\, ,\quad \mathcal{K}_{222}=3\,,
\end{equation}
and the second Chern class is
\begin{equation}
\int_{\vec{H}}c_2(X)=\begin{pmatrix}
52\\
42\\
24
\end{pmatrix}\, .
\end{equation}
In the limit $\xi^1\rightarrow 0$, keeping $\xi^{2,3}>0$, one approaches the wall in the K\"ahler cone that separates $\text{CY}_1$ from $\text{CY}_2$. The holomorphic curve class represented by $(1,0,0)$ is GV-nilpotent of order one, and lies outside the interior cone. Because this class is a generator of the Mori cone, it also lies outside the cone generated by all other curves with non-vanishing GV invariants. Thus, as we approach the wall of the K\"ahler cone only the instanton corrections in \eqref{eq:Finst} from the curve class $(1,0,0)$ become unsuppressed and have to be resummed using \eqref{eq:reflection_formula}. Only the period $\mathcal{F}_1$ develops a logarithmic singularity, indicating a conifold singularity in the mirror dual $\tilde{X}$.

For later reference (using the methods developed in \cite{Hosono:1993qy,Hosono:1994ax}) we compute all the non-vanishing GV invariants $n^0_{(i,j,k)}$ with $j+k\leq 2$ and arbitrary $i$,
\begin{equation}\label{eq:GVinvariants}
	\centering
	\begin{tabular}{l | l | l | l | l}
	$n^0_{(i,0,0)}$  & $n^0_{(i,1,0)}$ & $n^0_{(i,0,1)}$ & $n^0_{(i,2,0)}$ & $n^0_{(i,1,1)}$\\ \hline	
	$n^0_{(1,0,0)}=2$ & $n^0_{(0,1,0)}=252$  & $n^0_{(0,0,1)}=2$ & $n^0_{(0,2,0)}=-9252$  & $n^0_{(1,1,1)}=2376$\\
		              & $n^0_{(1,1,0)}=2376$ & $n^0_{(1,0,1)}=2$ & $n^0_{(1,2,0)}=10260$& $n^0_{(2,1,1)}=2376$\\
		              & $n^0_{(2,1,0)}=252$  &                   & $n^0_{(2,2,0)}=206712$&\\
		              & & & $n^0_{(3,2,0)}=10260$&\\
		              & & & $n^0_{(4,2,0)}=-9252$
	\end{tabular}
\end{equation}
Since $n^0_{(1,0,0)}=2$, we expect to find two conifold singularities in the corresponding limit in complex structure moduli space of $\tilde{X}$. We will confirm this in the next section.

We now choose an orientifold using the involution
\begin{equation}\label{caliis}
\mathcal{I}:\, x_2\mapsto -x_2\, .
\end{equation}
The fixed locus in the ambient variety is
\begin{align}
\mathfrak{F}_\mathcal{I}=&\{x_2=0\}\cup\{x_1=x_3=x_4=0\}\cup \{x_1=x_5=x_7=0\}\, .
\end{align}
The generic $\mathbb{Z}_2$-even anticanonical polynomial is non-vanishing along these loci.
The third locus does not intersect $X$, while the first two intersect $X$ transversally.
The first locus gives rise to an O7-plane on the divisor $D_2$, and the second gives rise to a single O3-plane.  Using the adjunction formula one computes $\chi_{D_2}=207$. Placing four D7-branes on top of the O7-plane, the D7-brane tadpole is canceled and the total induced D3-brane charge on the O7-plane plus the D3-brane charge of the O3-plane is
\begin{equation}
Q=-\frac{\chi_{D_2}}{4}-\frac{1}{4}=-\frac{207}{4}-\frac{1}{4}=-52\, .
\end{equation}
Therefore, we can turn on three-form fluxes $(F_3,H_3)$ with
\begin{equation}
Q_{D3}^{\text{flux}}:= \frac{1}{2}\int_X F_3\wedge H_3\leq 52\, .
\end{equation}
We have $h^{1,1}_-(X,\mathcal{I})=0$ because the toric divisors generate $H_4(X,\mathbb{Z})$ and they are invariant under the orientifold action. From the Lefschetz fixed point theorem one computes
\begin{equation}
h^{2,1}_-(X,\mathcal{I})=h^{1,1}_-(X,\mathcal{I})-Q-\frac{\chi_{CY}}{4}-1=0+52-(-48)-1=99=h^{2,1}(X)\, .
\end{equation}
Thus we have $h^{1,1}_-(X,\mathcal{I})=h^{2,1}_+(X,\mathcal{I})=0$. As a consequence, none of the moduli are projected out. In the following we will use the involution $\mathcal{I}$ to define an involution $\tilde{\mathcal{I}}$ in the mirror threefold $\tilde{X}$.

\subsection{The Greene-Plesser mirror dual}\label{ss:tildex}
Next, we construct the orbifold $X/G$. We start by computing the dual polytope $\Delta:= (\Delta^\circ)^\circ$, $\Delta\subset M\simeq \mathbb{Z}^4$.  Its vertices are the columns of
\begin{equation}
\begin{pmatrix}
1 & 1  & -5 & -11 & 1 & 1 \\
2 & 0 & -4 & -10 & 2 & 2 \\
0 & 0 & 0 & -6 & 3 & 0 \\
0 & 0 & -6 &  -6 & 0 & 6
\end{pmatrix}\, .\label{eqn:dual poly 0}
\end{equation}
We have an embedding
\begin{align}
\imath:\quad & N\hookrightarrow M\, ,\quad n\mapsto \Lambda n\, ,\quad  \Lambda=\begin{pmatrix}
11 & 10 & 6 & 6\\
10 & 10 & 6 & 6\\
6 & 6 & 3 & 6\\
6 & 6 & 6 & 0
\end{pmatrix}\, ,
\end{align}
which is a group homomorphism that maps the vertices of $\Delta^{\circ}$ to the vertices of $\Delta$. The Greene-Plesser group is the group coset $G:= N/\imath(M)\simeq \mathbb{Z}_6\times \mathbb{Z}_6$. The two $\mathbb{Z}_6$ factors can be chosen to act on the toric coordinates with charges
\begin{equation}
\vec{w}_1=(0,3,0,0,1,1,1)\, ,\quad \vec{w}_2=(0,0,0,0,5,0,1)\, .\label{eqn:GP actions}
\end{equation}
The points in $\Delta^\circ$ not interior to facets are mapped to the vertices of $\Delta$, and to the further point $(-5,-4,-3,0)^t$. These seven points correspond to seven $G$-invariant monomials of the anticanonical line bundle of our toric fourfold $Y$,
\begin{equation}\label{eq:polyLCS}
f(\vec{x})=\psi_0\prod_{i=1}^{7}x_i-\psi_1x_1^6 -\psi_2 x_2^2-\psi_3 x_4^6x_6^6x_7^6-\psi_4 x_3^6x_4^{12}x_7^6-\psi_5 x_5^3x_6^6x_7^3-\psi_6 x_3^6x_5^6-\psi_7 x_3^6x_4^6x_5^3x_7^3\, .
\end{equation}
This represents the \textit{generic} anticanonical polynomial defining $X/G$, or equivalently the generic $G$-invariant polynomial defining a symmetric Calabi-Yau $X$. There exists a special locus in moduli space where the $G$-symmetric $X$ develops a set of $18$ conifold singularities. To see this one considers the patch where $x_{3,5,7}\neq 0$, where we can gauge fix (part of) the toric scaling relations by setting $x_3=x_5=x_7=1$. Note that this leaves a residual scaling equivalence $(x_4,x_6)\sim (-x_4,-x_6)$, because the toric scaling relation associated with the third row of the GLSM charge matrix in \eqref{eqn:GLSM} preserves our gauge fixing condition for scaling parameters $\pm 1$. Furthermore, we use up the action of the algebraic torus on $X$ and the freedom to rescale $f$ to set $\psi_{0,2,3,4,5}=1$. Then, along a codimension-one locus in moduli space specified by
\begin{equation}\label{eq:conifold_locus}
\psi_7=1+\psi_6\, ,
\end{equation}
one finds that $f=df=0$ at the following set of points in $X$,
\begin{equation}\label{eq:conifold_position}
x_1=x_2=0\, ,\quad x_4^6=-1\, ,\quad x_6^6=1-\psi_6\, .
\end{equation}
Na\"ively this is a set of $6\times 6=36$ conifold singularities, but we need to account for the residual scaling equivalence $(x_4,x_6)\sim (-x_4,-x_6)$ which implies that there are only $18$ inequivalent conifolds.
One can show that these $18$ conifolds can be resolved to give the anticanonical hypersurface in the toric fourfold specified by the GLSM charge matrix
\begin{equation}
\begin{pmatrix}
	0 & -1 & 0 & 0 & 0 & 1 & 1 & 0\\
	0 & 0  & 1 & -1& 0 & 0 & -1 & 1\\
	1 & 3 & -1 & 1 & 0 & 0 & -1 & 0\\
	0 & 0 & 0 & 1 & 1 & 0 & 0 & -2
	\end{pmatrix}
\end{equation}
with positive FI parameters. The first row corresponds to the resolution $\mathbb{P}^1$ and indeed has GV invariant equal to $18$.

The gauge-invariant coordinates adapted to the LCS patch are $\tilde{\psi}_a=\prod_{i=1}^{7}(\psi_i)^{Q^a_i}$, i.e.~in our gauge we have
\begin{equation}
\tilde{\psi}_1=\psi_7\, ,\quad \tilde{\psi}_2=\psi_1\, ,\quad \tilde{\psi}_3=\frac{\psi_6}{\psi_7^2}\, ,
\end{equation}
and the flat LCS coordinates mirror-dual to curve volumes are
\begin{equation}
z^a=\frac{\ln(\tilde{\psi}_a)}{2\pi i}+\sum_{\vec{n}\in \mathbb{N}^3_{0}}\alpha^a_{\vec{n}}\prod_{b=1}^{3}\tilde{\psi}_b^{n_b}\, ,
\end{equation}
with coefficients $\alpha_{\vec{n}}^a$ that can be computed systematically as in \cite{Hosono:1994ax}. Let us define
\begin{equation}
\Psi(\tilde{\psi}_3):=\frac{1-\sqrt{1-4\tilde{\psi}_3}}{2\tilde{\psi}_3} = 1+\tilde{\psi}_3+\mathcal{O}(\tilde{\psi}_3^2)\, .
\end{equation}
In terms of the $\tilde{\psi}_a$, the conifold locus \eqref{eq:conifold_locus} occurs when $\tilde{\psi}_1=\Psi(\tilde{\psi}_3)$. In terms of the flat coordinates $z^a$
this simply corresponds to the locus $z^1= 0$, as follows from the identity
\begin{equation}
\ln\Bigl(\Psi(\tilde{\psi}_3)\Bigr)+2\pi i\sum_{\vec{n}\in \mathbb{N}^3_{0}}\alpha^1_{\vec{n}}\, \Psi(\tilde{\psi}_3)^{n_1}\tilde{\psi}_2^{n_2}\tilde{\psi}_3^{n_3}=0\,,
\end{equation}
which one may verify order by order in $\tilde{\psi}_{2}$, $\tilde{\psi}_{3}$.\footnote{We have verified this to order 42 in $\tilde{\psi}_{2,3}$.} Thus, we see that keeping $\text{Im}(z_{2,3})> 1$ while sending $z_1\rightarrow 0$ produces the 18 conifold singularities that we just analyzed.

Next, we take $\tilde{X}$ at its orbifold point in K\"ahler moduli space, and consider the induced action of $\mathcal{I}:X\rightarrow X\, ,[x_1:...:x_7]\mapsto [x_1:-x_2:x_3:...:x_7]$ on $\tilde{X}$,
\begin{equation}
\tilde{\mathcal{I}}: \, \tilde{X}\rightarrow \tilde{X}\, ,\quad \pi([x_1:...:x_7])\mapsto \pi([x_1:-x_2:...:x_7])\, ,
\end{equation}
defined to act on representatives $[x_1:...:x_7]$ as $\pi\circ\mathcal{I}$. Here, $\pi: X\rightarrow X/G$ is the projection mod $G$. Since $G$ commutes with $\mathcal{I}$, the involution $\tilde{\mathcal{I}}$ is well-defined. Clearly, no complex structure moduli are projected out by the orientifolding, i.e.~$h^{2,1}_+(\tilde{X},\tilde{\mathcal{I}})=0$.
This is simply because none of the complex structure moduli of $X$ were projected out by orientifolding by $\mathcal{I}$.  Furthermore, we can extend the action of $\tilde{\mathcal{I}}$ to the resolution of the orbifold in such a way that $h^{1,1}_-(\tilde{X},\tilde{\mathcal{I}})=0$. This is done by letting all toric coordinates associated with the resolution divisors transform trivially under the involution $\tilde{\mathcal{I}}$. This leaves the inherited divisor classes invariant, and one can show that in fact all divisor classes are invariant under the involution --- see Appendix \ref{App:1proveh11-=0}. Using this and the Lefschetz fixed point theorem one computes the D3-brane tadpole for
$\mathfrak{so}(8)$ stacks as
\begin{equation}\label{eq:Lefschetz}
-Q=\frac{\chi(\mathfrak{F}_{\tilde{\mathcal{I}}})}{4}=\frac{1}{2}\Bigl(h^{1,1}(\tilde{X})+h^{2,1}(\tilde{X})\Bigr)-\Bigl(h^{1,1}_-(\tilde{X},\tilde{\mathcal{I}})+h^{2,1}_+(\tilde{X},\tilde{\mathcal{I}})\Bigr)+1=52\,,
\end{equation}
where $\chi(\mathfrak{F}_{\tilde{\mathcal{I}}})$ denotes the Euler characteristic of the fixed locus $\mathfrak{F}_{\tilde{\mathcal{I}}}$ of the involution $\tilde{\mathcal{I}}$.
Alternatively, one can directly compute the Euler characteristic of the fixed locus in the orbifold limit (see Appendix \ref{App:2proveh11-=0}), which agrees with \eqref{eq:Lefschetz}.
Since $h^{1,1}_+(\tilde{X},\tilde{\mathcal{I}})=h^{1,1}(\tilde{X},\tilde{\mathcal{I}})$ we may go away from the orbifold point in K\"ahler moduli space without affecting the periods. As a consequence, we have defined an orientifold of the \textit{resolved} orbifold $\tilde{X}$.

It remains to show that the conifolds in $X$ correspond to conifolds in $\tilde{X}$, and to determine how many conifold singularities arise in the singular limit $z_1\rightarrow 0$. To do so, we will need to slightly change the gauge of the defining polynomial $f(\vec{x})$.
In eq.~\eqref{eq:polyLCS} we have used the continuous $G$-compatible ambient space automorphisms
\begin{equation}\label{eq:Gautomorphism}
x_2\mapsto x_2+\lambda x_1x_3x_4x_5x_6x_7\, ,\quad \lambda\in \mathbb{C}\, ,
\end{equation}
in order to eliminate the monomial $x_1^2x_3^2x_4^2x_5^2x_6^2x_7^2$. In order to analyze the orientifold, instead we would like to restrict to a manifestly orientifold-invariant polynomial $f$. Starting from \eqref{eq:polyLCS} in our gauge $\psi_0=\psi_2=1$, and using \eqref{eq:Gautomorphism} with $\lambda=-1/2$ amounts to replacing
\begin{equation}\label{eq:ourgt}
\prod_i x_i\mapsto  -\frac{1}{4}x_1^2x_3^2x_4^2x_5^2x_6^2x_7^2\, ,
\end{equation}
which makes the defining polynomial $f$ manifestly invariant under $\tilde{\mathcal{I}}$.

Since \eqref{eq:Gautomorphism} acts trivially on the locus $x_1=x_2=0$ where the conifolds reside, their position is not altered by \eqref{eq:ourgt}.  In $X$, the curve $x_1=x_2=0$ can be shown (see Appendix \ref{App:OrbifoldEuler}) to be a fixed curve of a $\mathbb{Z}_2$ subgroup of the Greene-Plesser group $G=\mathbb{Z}_6\times \mathbb{Z}_6$. The coset $G/\mathbb{Z}_2$ acts transitively on the $18$ conifolds, so after modding out by $G/\mathbb{Z}_2$ we retain only a single conifold singularity. Finally, we have to mod out by the remaining $\mathbb{Z}_2$ symmetry that maps the curve $x_1=x_2=0$ to itself pointwise. Locally, around a solution of \eqref{eq:conifold_position}, we may embed the conifold in $\mathbb{C}^4\ni (x,y,u,v)$ via the vanishing of the polynomial
\begin{equation}
P(x,y,u,v)=x^2+y^2+u^2+v^2-\epsilon+...=0\, ,
\end{equation}
with deformation parameter $\epsilon$ such that we have $dP=P=0$ on the locus $x=y=u=v=0$ in the singular limit $\epsilon\rightarrow 0$. Here,
\begin{equation}
x:= \frac{x_6^0}{2}x_1\, ,\quad y:= ix_2\, ,\quad \begin{pmatrix}
u\\
v
\end{pmatrix}:=\begin{pmatrix}
-6i & 3i(1-\psi_6)\\
0   & 3(1-\psi_6)
\end{pmatrix}\begin{pmatrix}
\tfrac{x_4}{x_4^0}-1\\
\tfrac{x_6}{x_6^0}-1
\end{pmatrix}\, ,
\end{equation}
and $(x_4^0,x_6^0)$ is a solution to eq.~\eqref{eq:conifold_position}. Here, $\epsilon:= 1+\psi_6-\psi_7$, and we neglect higher-order corrections in $(x,y,u,v)$ as well as non-constant terms that vanish in the limit $\epsilon\rightarrow 0$.

Locally, the $\mathbb{Z}_2$-orbifold action is given by
\begin{equation}\label{eq:Z2orbifold}
\mathbb{Z}_2:\quad (x,y)\mapsto (-x,-y)\, ,
\end{equation}
and the local action of the orientifold involution is
\begin{equation}\label{eq:local_involution}
\tilde{\mathcal{I}}:\quad (x,y)\mapsto (x,-y)\overset{\mathbb{Z}_2}{\sim} (-x,y)\, ,
\end{equation}
so there is an O7-plane on the divisor $x=0$ as well as on $y=0$.

\begin{figure}
	\centering
	\begin{tabular}{p{5cm} p{5cm} p{5cm}}
		\includegraphics[keepaspectratio,height=5cm]{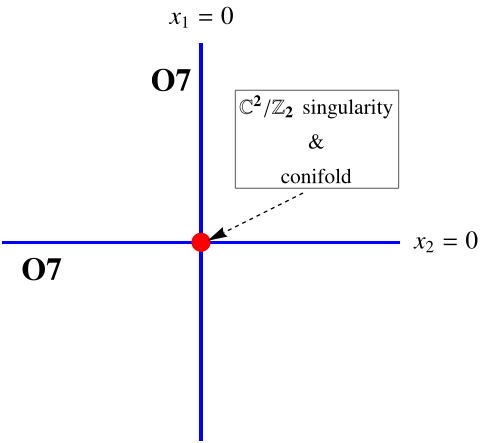} &
		\includegraphics[keepaspectratio,height=5cm]{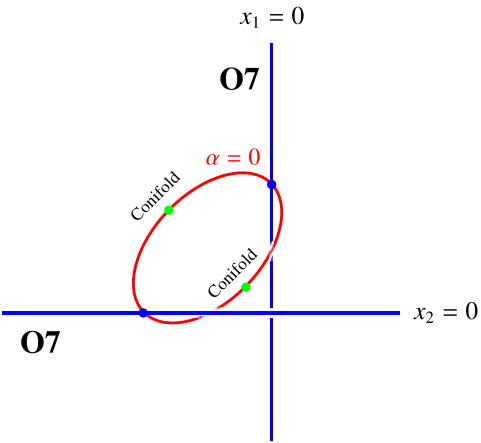} &
		\includegraphics[keepaspectratio,height=5cm]{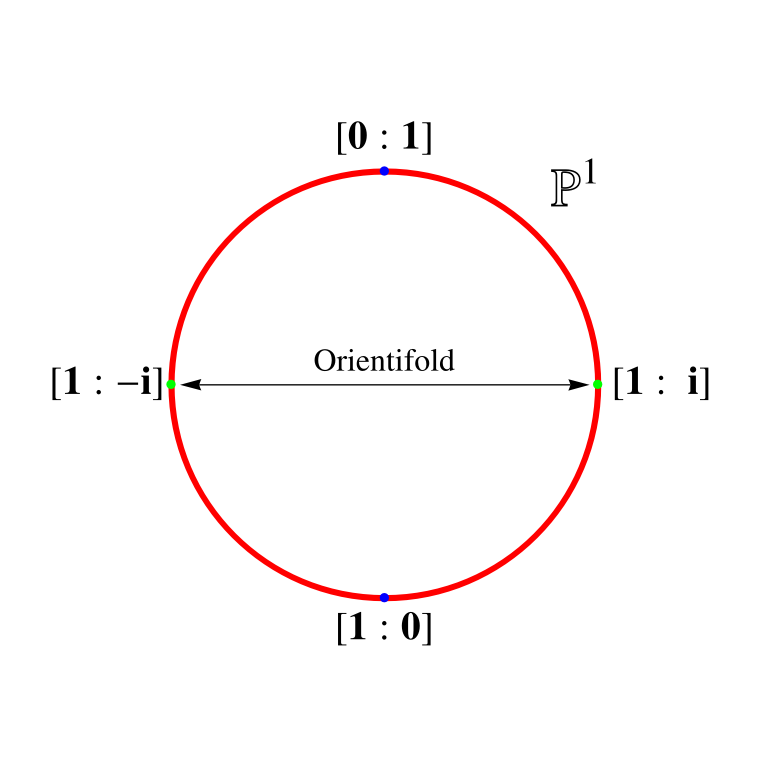}
	\end{tabular}

	\caption{Left: the slice $\{u=v=0\}=\mathbb{C}^2/\mathbb{Z}_2$ and the position of the orientifold planes $x_1=0$ and $x_2=0$. Two O7-planes intersect at the orbifold singularity $x_1=x_2=0$ which also contains the conifold singularity. Middle and right: the same slice after the resolution of the orbifold singularity and a closeup of the exceptional divisor. The O7-planes intersect the exceptional divisor $\alpha=0$ at antipodal points $[1:0]$ and $[0:1]$. The conifold singularities reside at $[1:i]$ and $[1:-i]$ and are mapped into each other by the orientifold involution.}
	\label{fig:singular_orientifold}
\end{figure}

\begin{figure}
	\centering
	\includegraphics[keepaspectratio,height=5cm]{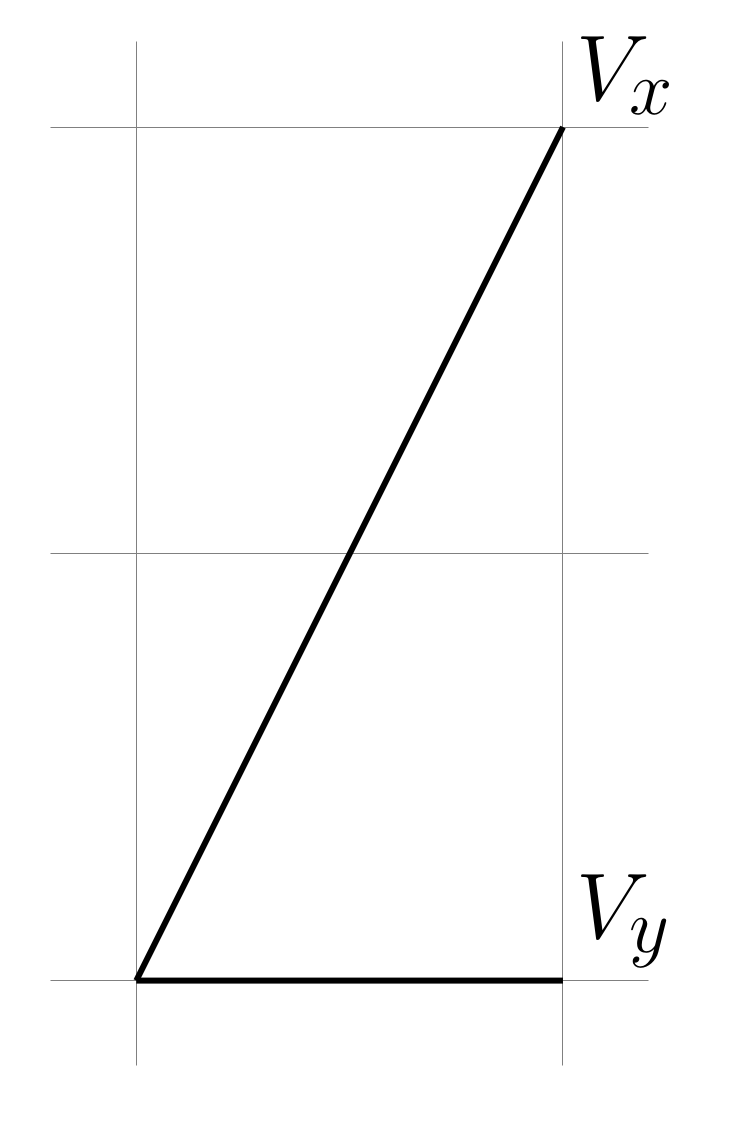}$\quad \quad\quad\quad$
	\includegraphics[keepaspectratio,height=5cm]{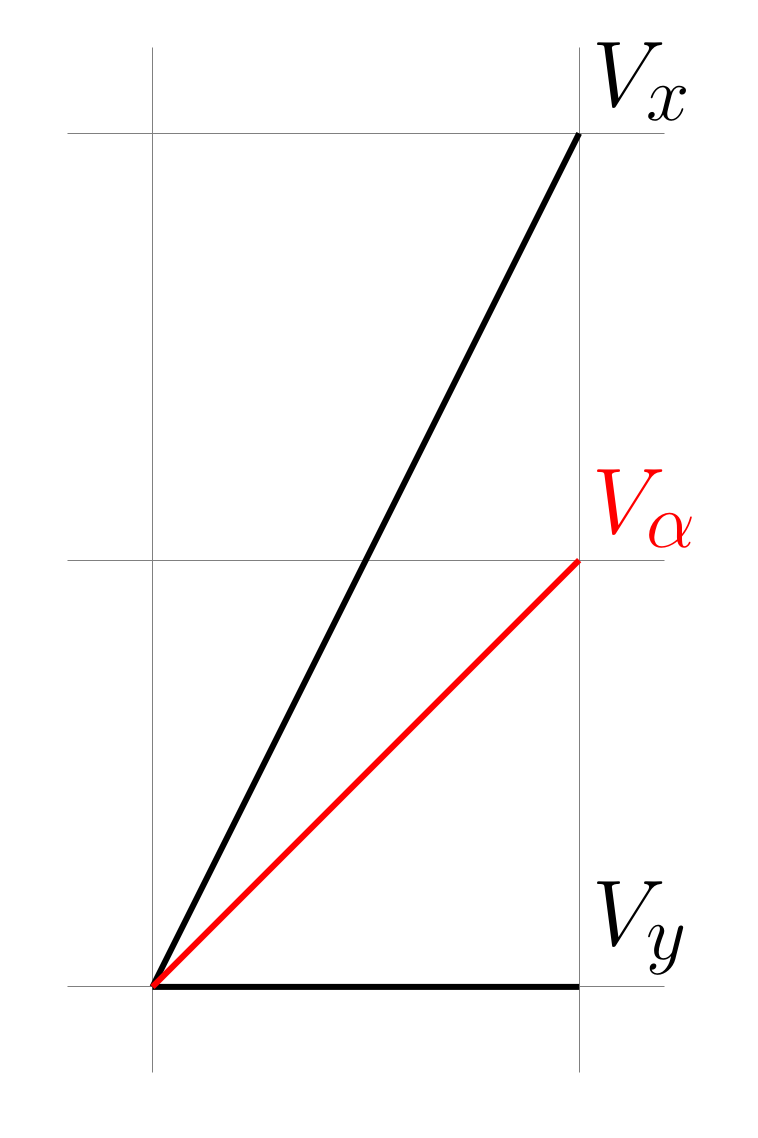}
	\caption{Left: the toric fan of the singular surface $\mathbb{C}^2/\mathbb{Z}_2$ with with two vertices $v_x=(1,2)$ and $v_y=(1,0)$. Right: the toric fan of the resolution of $\mathbb{C}^2/\mathbb{Z}_2$ by a resolution divisor $\alpha=0$ associated with the vertex $v_{\alpha}=\frac{1}{2}(v_x+v_y)=(1,1)$.}
	\label{fig:C2modZ2}
\end{figure}

Orbifolding by \eqref{eq:Z2orbifold} produces an $A_1$ singularity in the ambient $\mathbb{C}^2\subset \mathbb{C}^4$ parameterized by $(x,y)$, and both orientifold planes and the conifold intersect at the singular locus (see Figure \ref{fig:singular_orientifold}).
As usual, we can resolve this singularity using toric geometry by introducing a blowup coordinate $\alpha$ (see Figure \ref{fig:C2modZ2} for the toric fan),
and a $\mathbb{C}^*$-scaling relation
\begin{equation}
(x,\alpha,y)\sim (\lambda x, \lambda^{-2}\alpha,\lambda y)\, ,\quad \lambda\in \mathbb{C}^*\, .
\end{equation}
The locus $x=y=0$ is removed (it is in the SR ideal of the toric fourfold) and replaced by the exceptional divisor $\alpha=0$. The polynomial $P$ is replaced by
\begin{equation}
\hat{P}(\alpha,x,y,u,v)=\alpha(x^2+y^2)+u^2+v^2-\epsilon=0\, ,
\end{equation}
In the limit $\epsilon\rightarrow 0$ we get not one but two conifold singularities. To see this, consider the exceptional divisor $\{\alpha=0\}\simeq \mathbb{P}^1\times \mathbb{C}^2$ parameterized by homogeneous coordinates $[x:y]\in \mathbb{P}^1$ and $(u,v)\in \mathbb{C}^2$. For $\epsilon=0$ we have $\hat{P}=d\hat{P}=0$ at the two points
\begin{equation}
\alpha=u=v=0\, ,\quad [x:y]=[1:\pm i]\, .
\end{equation}
Indeed, the GV invariant of the curve class $(1,0,0)$ that is mirror dual to the conifold $S^3$ has $n^0_{(1,0,0)}=2$, see eq.~\eqref{eq:GVinvariants}. Finally, the O7-planes at $x=0$ and $y=0$ intersect the exceptional divisor $\alpha=0$ at $[x:y]=[0:1]$ and $[x:y]=[1:0]$, respectively, so the two orientifold planes no longer intersect each other and the two conifold singularities are also separated from the O7-planes at finite blowup volume. The conifolds are moreover mapped into each other by the involution in eq.~\eqref{eq:local_involution} and are therefore identified in the orientifold. We conclude that in the limit $z^1\rightarrow 0$ there appears a single conifold at generic position in the orientifold of $\tilde{X}$.
\subsection{Explicit flux vacua}\label{sec:explicit}
Now we are ready to find explicit conifold vacua with small flux superpotential in the complex structure moduli space of the Calabi-Yau orientifold discussed in the previous section. We have $h^{1,1}_+(\tilde{X},\tilde{\mathcal{I}})=h^{1,1}(\tilde{X})=99$, and $h^{2,1}_-(\tilde{X},\tilde{\mathcal{I}})=h^{2,1}(\tilde{X})=3$. Thus, all three-form classes are orientifold-odd and we can turn on generic three-form fluxes $\vec{f},\vec{h}\in \mathbb{Z}^{8}$ on the three-cycles in $\tilde{X}$, compatible with the D3-brane tadpole bound
\begin{equation}
\frac{1}{2}\int_{\tilde{X}}F_3\wedge H_3=\frac{1}{2}{\vec{f}\,}^t \Sigma  \vec{h}\leq 52\, .
\end{equation}
We will search for appropriate flux quanta systematically as follows. As a first step, we compile a list of restricted candidate flux integers in a box:
\begin{align}
V_M:= &\Biggl\{\vec{M}\in \mathbb{Z}^3\Biggl.\Biggr|~0\neq M^a\in \{-200,200\} \,\,\forall a=1,2,3\, ,\quad M^3>0\, ,\quad  \det{\left(\sum_{a=1}^3\mathcal{K}_{ija}M^a\right)}\neq 0\,,\Biggr. \nonumber\\  &\quad \Biggl.\sum_{a=1}^3M^a\tilde{b}_a\in \mathbb{Z}\, ,\quad \sum_{a=1}^3a_{ia}M^a\in \mathbb{Z}\,\,\forall i=2,3\Biggr\}\, ,
\end{align}
with
\begin{equation}
a_{ia}=\begin{pmatrix}
2 & \frac{3}{2} & 0\\
0 & 0 & 0
\end{pmatrix}_{ia}\, ,\quad \tilde{b}_a=\frac{1}{24}\left(\begin{pmatrix}
52\\ 42\\ 24
\end{pmatrix}+\begin{pmatrix}
2\\
0\\
0
\end{pmatrix}\right)_a=\begin{pmatrix}
\frac{9}{4}\\
\frac{7}{4}\\
1
\end{pmatrix}_a\, .
\end{equation}
Such choices of $\vec{M}$ give rise to integer RR fluxes as in eq.~\eqref{eq:restricted_fluxes} that satisfy eq.~\eqref{eq:quantization_condition}, and we have gauge-fixed the center of $SL(2,\mathbb{Z})$ by enforcing that $M^3>0$.

For each element $\vec{M}\in V_M$, let $V_K^{(\vec{M})}$ be the set of integers $\vec{K}\in \mathbb{Z}^3$ that satisfy eq.~\eqref{eq:flat_Kfluxes} for the given choice of $\vec{M}$. Enumerating these involves solving a homogeneous Diophantine equation of degree \textit{two} in the variables $(K_2,K_3)\in \mathbb{Z}^2$ subject to the constraints $0\leq -\vec{M}^t \vec{K}\leq 104$ and $0<p^i<1$ with $p^i:= (\mathcal{K}_{ija}M^a)^{-1}K_j$. This can be done efficiently using \texttt{Mathematica}.\footnote{Note that quadratic Diophantine equations are solvable, in contrast to the generic case \cite{siegel1972theorie,mathematical1987encyclopedic}.} The resulting set of flux integers in
\begin{equation}
V:= \Bigl\{(\vec{M},\vec{K})\in V_M\times V_K^{(\vec{M})}\Bigr\}
\end{equation}
give rise to perturbatively flat vacua of the superpotential $W_{\text{poly}}^{(0)}(z^i,\tau)$, i.e.~$W_{\text{poly}}^{(0)}(z^i,\tau)=dW_{\text{poly}}^{(0)}(z^i,\tau)=0$ along the loci where $z^i=p^i\tau$, compatible with the tadpole bound.

For each element of $V$, we stabilize the remaining light direction using the truncation of $W^{(0)}_{\text{inst}}$ in \eqref{eq:superpotential_split} to leading order in the instanton expansion, i.e.~we approximate
\begin{equation}\label{eq:Winst_racetrack}
W^{(0)}_{\text{inst}}(\tau)\approx A_{(1,0)} e^{2\pi i p^2\tau}+A_{(0,1)}e^{2\pi i p^3\tau}\, ,
\end{equation}
with
\begin{align}
A_{(1,0)}&:= \frac{1}{(2\pi i)^2}\sum_j (jM^1+M^2)n^0_{(i,1,0)}=\frac{2880(M^1+M^2)}{(2\pi i)^2}\, ,\\
A_{(0,1)}&:= \frac{1}{(2\pi i)^2}\sum_j (jM^1+M^3)n^0_{(i,0,1)}=\frac{2(M^1+2M^3)}{(2\pi i)^2}\, .
\end{align}
The above superpotential has $\del_{\tau}W^{(0)}_{\text{inst}}=0$ for
\begin{equation}\label{eq:dilatonvev}
e^{2\pi i \tau}=\left(-\frac{A_{(0,1)}p^3}{A_{(1,0)}p^2}\right)^{\frac{1}{p^2-p^3}}\, .
\end{equation}
Generically, we have $A_{(1,0)}\gg A_{(0,1)}$, so one stabilizes at weak coupling, $\text{Im}(\tau)>1$, if $p^2>p^3$. If furthermore $p^2\approx p^3$, then $e^{2\pi i \tau}$ is in fact exponentially small.

Of course, the true F-term equations also contain the K\"ahler covariantization of the partial derivative $\del_\tau\rightarrow D_{\tau}=\del_{\tau}+\del_{\tau}K_{\text{eff}}$, with effective K\"ahler potential obtained by evaluating \eqref{eq:KahlerCS} along the flat valley, i.e.
\begin{equation}
K_{\text{eff}}(\tau,\bar{\tau})=-4\ln\Bigl(-i(\tau-\bar{\tau})\Bigr)+\mathcal{O}\Bigl(\text{Im}(\tau)^{-3}\Bigr)+\text{constant}\, .
\end{equation}
For large $\text{Im}(\tau)$ this gives a small correction to eq.~\eqref{eq:dilatonvev}, and even for relatively small values one still finds nearby vacua of the actual F-term equation.

\begin{figure}
	\centering
	\includegraphics[keepaspectratio,width=7.5cm]{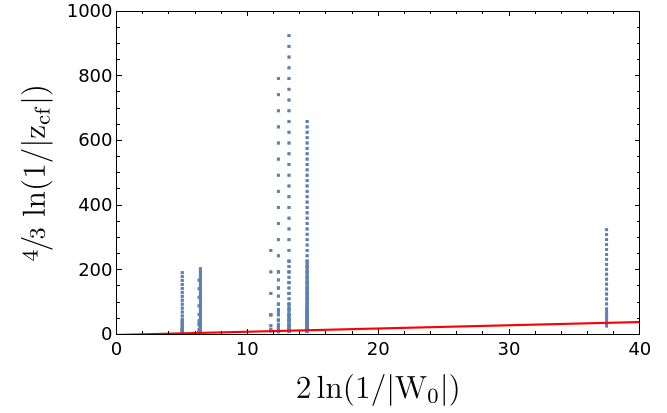}
	\includegraphics[keepaspectratio,width=7.5cm]{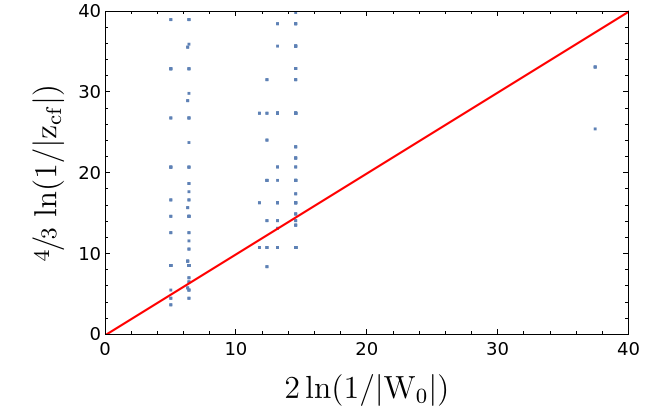}
	\caption{Scatter plots showing the values of $2\ln(1/|W_0|)$ and $\frac{4}{3}\ln(1/|\zc|)$ with diagonal in red indicating the critical region where the uplift potential of an anti-D3-brane would compete efficiently with KKLT bulk moduli stabilization. Left: All vacua. Right: Only vacua that live close to the critical line.}
	\label{fig:scatterplot_sugra_all}
\end{figure}

\begin{figure}
	\centering
	\includegraphics[keepaspectratio,width=8cm]{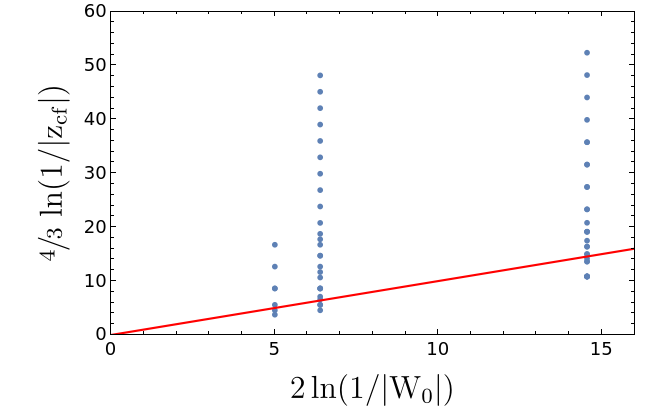}
	\caption{A scatter plot as in Figure \ref{fig:scatterplot_sugra_all} but showing only vacua with $g_sM>1$.}
	\label{fig:scatterplot_sugra}
\end{figure}

We will consider only those fluxes in $V$ for which the next-to-leading corrections to $W^{(0)}_{\text{inst}}$ are suppressed at least at the $10\%$ level relative to the leading terms in \eqref{eq:Winst_racetrack}, and for which $|\zc|\ll |W_0|\ll 1$. This leaves us with 696 vacua, for which
we show the values of $|W_0|$ and $|\zc|$ in Figure \ref{fig:scatterplot_sugra_all}.\footnote{Strictly speaking, each of these vacua again comes as a family because we have not specified the flux integer $P$ which can be freely chosen in a fundamental domain of the conifold monodromy $0\leq P< |M|$. Since it only affects the phase of the conifold modulus, and does not contribute to the D3-brane charge its value is of no relevance to us.}
For most of these the Klebanov-Strassler sector has small 't Hooft coupling $|g_sM|<1$, and the smallest value of $W_0$ that we find is given by (see Table \ref{tab:vacua})
\begin{equation}
\min |W_0|\approx 7.4\times 10^{-9}\, .
\end{equation}
However, there are also 63 vacua with $|g_sM|>1$
that may somewhat marginally live in the ten-dimensional supergravity regime.\footnote{Although it is well-known that $g_s M \gg 1$ is sufficient for the infrared region of the throat to be weakly curved in string units, we are not aware of a specific numerical threshold $(g_s M)_{\text{min}}$ that demarcates the region below which the supergravity approximation fails.  Determining such a threshold would be worthwhile.}
We find a maximum value (see Table \ref{tab:vacua})
\begin{equation}
\max |g_sM|\approx 2.8\, .
\end{equation}
The 63 vacua with potential supergravity throats come in three families with flux superpotentials $|W_0|\approx \left\{6.9\times 10^{-4},4.1 \times 10^{-2},8.1 \times 10^{-2}\right\}$, but with vastly different values of $|\zc|$. We have checked that the neglected two-instanton corrections are suppressed by a relative factor of order $|W_0|$ and the three-instanton corrections are suppressed by further such factors. We show the values of $|W_0|$ and $|\zc|$ in Figure \ref{fig:scatterplot_sugra}.

Let us walk through the stabilization steps for one of these vacua. We consider $\vec{M}=(4,-8,8)$ and $\vec{K}=(-8,3,-6)$. These correspond to the family of flux vectors
\begin{equation}
\vec{f}=(3,P,-4,0,0,4,-8,8)^t\, ,\quad \vec{h}=(0,-8,3,-6,0,0,0,0)^t\, ,
\end{equation}
and we may choose $P=0$. We have
\begin{equation}
Q^{D3}_{\text{flux}}=\frac{1}{2}{\vec{f}\,}^t\Sigma \vec{h}=-\frac{1}{2}\vec{M}^t\vec{K}=52\, ,
\end{equation}
so we exactly saturate the tadpole bound. Furthermore,
\begin{equation}
N_{ij}:= \sum_a\mathcal{K}_{ija}M^a=8\begin{pmatrix}
1& -1 \\ -1 & 0
\end{pmatrix}_{ij}\, ,\quad p^i:= (N^{-1})^{ij}K_j=\frac{3}{8}\begin{pmatrix}
2\\
1
\end{pmatrix}^i\, ,\quad p^iK_i=0\, .
\end{equation}
Thus, we stabilize the two bulk moduli $z^{2,3}$ and the dilaton along the flat valley
\begin{equation}
z^i=\frac{3}{8}\tau\begin{pmatrix}
2\\
1
\end{pmatrix}^i\, .
\end{equation}
In the basis $(\tau,z^2,z^3)$, the matrix of second derivatives of the superpotential $W_{\text{poly}}^{(0)}(\tau,z^i)$ reads
\begin{equation}
\del^2W_{\text{poly}}^{(0)}(\tau,z^i)=\begin{pmatrix}
0 & -K_j\\
-K_i & N_{ij}
\end{pmatrix}=\begin{pmatrix}
0  & -3 & 6\\
-3  &  8 &-8\\
6  & -8 & 0
\end{pmatrix}\, ,
\end{equation}
which has eigenvalues $(0,4\pm 5\sqrt{5})$. Thus, we may indeed integrate out the field directions away from the flat valley, and consider the effective theory along the valley in the next step. The effective superpotential $W^{(0)}_{\text{inst}}$ takes the form
\begin{equation}
W_{\text{inst}}^{(0)}(\tau) \approx A_{(1,0)} e^{2\pi i \frac{3}{4}\tau}+A_{(0,1)}e^{2\pi i \frac{3}{8}\tau}\, ,
\end{equation}
with
\begin{align}
A_{(1,0)}&=\frac{2880(M^1+M^2)}{(2\pi i)^2}=\frac{-11520}{(2\pi i)^2}\, ,\quad
A_{(0,1)}=\frac{2(M^1+2M^3)}{(2\pi i)^2}=\frac{40}{(2\pi i)^2}\, ,
\end{align}
so we stabilize $\tau$ near
\begin{equation}
e^{2\pi i \tau_0}=\left(\frac{40\cdot 3/8}{11520 \cdot 3/4}\right)^{\frac{8}{3}}\approx 4.4\times 10^{-8}\, .
\end{equation}
Replacing $\del_{\tau}\rightarrow D_\tau$ shifts this slightly to $g_s\approx 0.38$. Furthermore, the vacuum value of the superpotential is
\begin{equation}
|W_0|\approx \sqrt{\tfrac{2}{\pi}} \,\Bigl|W^{(0)}_{\text{inst}}(\tau_0)\Bigr|\approx 6.9\times 10^{-4}\, ,
\end{equation}
neglecting the contribution from $W^{(1)}\zc$. The neglected two-instanton corrections are suppressed by a relative $\mathcal{O}(10^{-3})$ factor and the three-instanton corrections are suppressed by a further $\mathcal{O}(10^{-3})$ factor, so the value of $|W_0|$ is a good measure of control of the instanton expansion.

In the final step, we stabilize the conifold modulus $\zc$ with the superpotential $W_{\text{cf}}(\zc)$. The conifold fluxes are
\begin{equation}
K':= K_1-M^a\mathcal{K}_{1ai}p^i=-5\, ,\quad M:= -M^1=-4\, ,
\end{equation}
so we stabilize the conifold modulus at
\begin{equation}
|\zc|=\frac{1}{2\pi}e^{-2\pi\frac{K'}{2g_sM}}\approx 5 \times 10^{-6}\, .
\end{equation}
The Klebanov-Strassler theory is (marginally) in its supergravity regime with somewhat large 't Hooft coupling $g_sM\approx 1.5$, and the infrared warp factor is of order
\begin{equation}
e^{2A}|_{\text{min}}\sim |\zc|^{\frac{2}{3}}\approx 2.9\times 10^{-4} \, .
\end{equation}
We note that $e^{2A}|_{\text{min}}$ and $|W_0|$ are of the same order, as one would want for a KKLT uplift from including an anti-D3-brane in the warped region. In Table \ref{tab:vacua} we list some interesting flux vacua.
\begin{table}
	\centering
	\begin{tabular}{c c | c  c c c c}
		$\vec{M}$ & $\vec{K}$ & $|W_0|$ & $|\zc|$ & $ \frac{|\zc|^{2/3}-|W_0|}{|W_0|}$ & $g_sM$ & $\eps$ \\ \hline
		
		$({4, -8, 10})$ & $({-6, 3, -4})$ & $7.4\times 10^{-9}$ &$5.4\times 10^{-14}$ & $-0.8$ & $0.6$ & $\sim 10^{-7}$\\
		
		$({8, -12, 6})$ & $({-5, 1, -2})$ & $6.9\times 10^{-4}$ & $1.4\times 10^{-5}$ & $-0.2$ & $1.0$ & $\sim 10^{-3}$\\
		
		$({-8, 4, 12})$ & $({5, 1, -4})$ & $4.1\times 10^{-2}$ & $5.2\times 10^{-3}$ & $-0.3$ & $2.8$ & $4\times 10^{-2}$\\
		
		$({-14, 6, 27})$ & $({4, 1, -2})$ & $1.4\times 10^{-3}$ & $5.3\times 10^{-5}$ & $\phantom{-} 0.03$ & $0.9$ & $\sim 10^{-3}$
	
	\end{tabular}
\caption{Some interesting vacua. First line: smallest $W_0$. Second line: smallest $W_0$ with $g_sM>1$.  Third line: largest $g_sM$. Fourth line: best alignment between $\zc^{2/3}$ and $W_0$. The parameter $\eps$ is the magnitude of the neglected two-instanton contributions to the superpotential relative to the retained one-instanton terms.}
\label{tab:vacua}
\end{table}

\newpage
\section{Discussion}\label{sec:conclusions}

In this work we have demonstrated that the mechanism of \cite{Demirtas:2019sip} for constructing flux vacua of type IIB string theory with exponentially small values of the flux superpotential can be applied not just at large complex structure, as in \cite{Demirtas:2019sip}, but also near conifold points in moduli space.  We laid out a procedure for finding \emph{conifold vacua} in which the flux superpotential is small.

The key challenge was to compute, and then to cancel, an order-one contribution to the superpotential coming from flux on the conifold cycles.  To accomplish this we considered the case in which the shrinking three-cycle of the conifold in a Calabi-Yau $\tilde{X}$ is mirror to a shrinking curve in the mirror threefold $X$.  Computing the prepotential for the complex structure moduli space of $\tilde{X}$, and then resumming the terms corresponding to type IIA worldsheet instantons wrapping the shrinking curve in $X$, we obtained the flux superpotential for type IIB compactification on $\tilde{X}$, including the term resulting from fluxes on the shrinking three-cycle of the conifold.  We then applied the mechanism of \cite{Demirtas:2019sip} to find fluxes for which the total flux superpotential, including the conifold term, is exponentially small.

We illustrated our approach in flux compactification of type IIB string theory on an orientifold of a Calabi-Yau threefold $\tilde{X}$ with $h^{1,1}(\tilde{X})=99$ and $h^{2,1}(\tilde{X})=3$.  To analyze $\tilde{X}$ and its orientifold we made heavy use of the Greene-Plesser construction: $\tilde{X}$ is the resolution of the orbifold $X/G$, with $X$ the mirror of $\tilde{X}$, and $G\simeq \mathbb{Z}_6\times \mathbb{Z}_6$ the Greene-Plesser group of $X$.  We found an O3/O7 orientifold involution $\tilde{\mathcal{I}}$ of $\tilde{X}$ leading to a D3-brane tadpole $-Q=52$, allowing reasonable freedom in choosing fluxes. We found many flux vacua and laid out in detail a flux choice for which $|W_0| \approx 7 \times 10^{-4}$, the conifold modulus $z_{\text{cf}}$ is stabilized at $|z_{\text{cf}}| \approx 5 \times 10^{-6}$, and the Klebanov-Strassler throat has $g_s M \approx 1.5$.

As classical flux vacua, our examples are rather well-controlled.  However, they are just a first step toward finding parametrically large Klebanov-Strassler throats in vacua with small values of the flux superpotential, and subsequently achieving K\"ahler moduli stabilization and a metastable uplift to de Sitter space.
Indeed, in our examples ten-dimensional supergravity is at best marginally valid near the tip of the throat, and at the same time the significant number of K\"ahler moduli makes stabilization computationally challenging.
We believe that these limitations have no deep relationship to our mechanism, but are instead accidental properties of the examples.  After all, we would expect to have to search through many candidate geometries to find one in which the flux superpotential is exponentially small, a throat region is parametrically large, and the numbers of moduli are small enough for simultaneous computational control of the geometry and its mirror.
Here we have examined one particularly tractable orientifold with $h^{2,1}(\tilde{X})=3$, and already finding therein a foundation for a parametrically controlled KKLT de Sitter vacuum would have been surprising to us.

One further limitation, however, is intrinsic to our mechanism, and will hold in all examples: at least one linear combination of the string coupling and the complex structure moduli remains rather light, with a mass of the same order as that of the K\"ahler moduli.  As noted in \cite{Demirtas:2019sip}, this is a departure from the original scenario of \cite{Kachru:2003aw}.  In the present context of conifold vacua, an important consequence is that the warp factor at the tip of the Klebanov-Strassler throat is set by the vev of a relatively light field.\footnote{Qualitatively similar results have been emphasized in \cite{Blumenhagen:2019qcg,Bena:2018fqc}, though the details and the causes are very different here.}
Metastable supersymmetry breaking in the presence of such light moduli will require further analysis.

Systematically enumerating conifold vacua in a much larger class of geometries will be very informative, but we leave this for future work.

\section*{Acknowledgements}

We thank Federico Carta, Naomi Gendler, Shamit Kachru, Cody Long, Daniel Longenecker, Andres Rios-Tascon, and Alexander Westphal for useful discussions.  We coordinated with the authors of the related work \cite{Alvarez-Garcia:2020pxd} so that our papers would appear simultaneously. The work of M.D., M.K., and L.M.~was supported in part by NSF grant PHY-1719877, and the work of L.M.~and J.M.~was supported in part by the Simons Foundation Origins of the Universe Initiative.

\appendix
\section{The D3-brane tadpole in the orbifold}\label{App:OrbifoldEuler}
In this appendix we will compute the Euler characteristic $\chi(\mathfrak{F}_{\tilde{\mathcal{I}}})$ of the fixed locus $\mathfrak{F}_{\tilde{\mathcal{I}}}$ of the orientifold involution $\tilde{\mathcal{I}}$, as promised in eq.~\eqref{eq:Lefschetz},
in two different ways. First, in \S\ref{App:1proveh11-=0} we will show directly that $h^{1,1}_-(\tilde{X},\tilde{\mathcal{I}})=h^{2,1}_+(\tilde{X},\tilde{\mathcal{I}})=0$ for the involution $\tilde{\mathcal{I}}:\tilde{X}\rightarrow \tilde{X}$, by using the description of $\tilde{X}$ as a smooth Calabi-Yau hypersurface in a toric fourfold $\tilde{Y}$. We then use the Lefschetz fixed point theorem to compute the Euler characteristic. Then, in \S\ref{App:2proveh11-=0} we go to the orbifold limit $\tilde{X}\rightarrow X/G$ and compute the Euler characteristic directly. The two computations agree.
\subsection{Computation in the resolved orbifold}\label{App:1proveh11-=0}
In this section, we verify that $h_{-}^{1,1}(\tilde{X},\tilde{\mathcal{I}})=0$ and $h_{-}^{2,1}(\tilde{X},\tilde{\mathcal{I}})=3$ under the orientifold involution $\tilde{\mathcal{I}}$ in the resolved orbifold.

First, let us briefly review how the anticanonical monomials and the automorphism group of the ambient fourfold $\tilde{Y}$ are determined from polytope data. Let $\Delta^{\circ}\in M$ be the Newton polyhedron for the anticanonical class of $Y$ and let $\Delta\in N$ be the dual polytope of $\Delta^{\circ}.$  A point $\rho\in\Delta\cap N$ corresponds to an edge of the toric fan of $Y$ and thus corresponds to a homogeneous coordinate $x_\rho.$ Similarly, each point $v\in\Delta^{\circ}\cap M$ determines a monomial $x^{\bm{v}}$,
\begin{equation}
x^{\bm{v}}=\prod_{\rho\in \Delta\cap N} x_\rho^{\langle v,\rho\rangle +1}\,.
\end{equation}
As was shown in \cite{Batyrev:1994hm}, a point $m\in \Delta^{\circ}\cap M$ strictly interior to a facet corresponds to a non-trivial so-called \textit{root automorphism}. Because $m$ is in a facet, there is a point $\rho_m\in \Delta\cap N$ such that $\langle m,\rho_m\rangle=-1$ and $\langle m,\rho\rangle >-1,$ $\forall \rho\neq \rho_m.$ Then the action of the automorphism of the group element $m$ is
\begin{equation}\label{eqn:root aut action}
x_{\rho_m}\mapsto x_{\rho_m}+\lambda_m \prod_{\rho'\neq \rho_m} x_{\rho'}^{\langle m,\rho'\rangle}\, ,\quad \lambda_m\in \mathbb{C}\, .
\end{equation}
Let $\mathcal{A}$ be the automorphism group of $\tilde{Y}$. Then, its connected component containing the identity is generated by the root automorphisms, as well as the action of the algebraic torus $(\mathbb{C}^*)^4\times \tilde{Y}\rightarrow \tilde{Y}$. Thus, the dimension of the automorphism group $\mathcal{A}$ is given by
\begin{equation}
\text{dim}\,\mathcal{A}=4+\sum_{\text{codim}\Theta^{\circ}=1}\ell^*(\Theta^{\circ})\,,
\end{equation}
where for a face $\Theta^{\circ} \subset \Delta^{\circ}$, $\ell^*(\Theta^{\circ})$ denotes the number of points in the interior of $\Theta^{\circ}$.

Given the anticanonical monomials and the automorphisms of $Y$, we can compute the number of K\"ahler moduli $h^{1,1}(\tilde{X})$ and the number of complex structure moduli $h^{2,1}(\tilde{X})$ of the Calabi-Yau hypersurface $\tilde{X}$. If a K\"ahler modulus is inherited from the ambient variety, then we call that K\"ahler modulus toric. We define $h_{\text{toric}}^{1,1}(\tilde{X})$ to be the number of toric K\"ahler moduli. Similarly, we define toric complex structure deformations to be deformations of the coefficients of the anticanonical monomials modulo the deformations that can be undone by  elements of $\mathcal{A}$, and modulo the overall scale. Likewise, we define $h_{\text{toric}}^{2,1}(\tilde{X})$ as the number of toric complex structure moduli.
We have
\begin{align}
h_{\text{toric}}^{2,1}(\tilde{X})=&~\#(\text{monomials})-\text{dim}\,\mathcal{A}-1=\sum_{\text{codim}\Theta^\circ\geq 2} \ell^*(\Theta^\circ)-4\,.
\end{align}
To determine $h_{\text{toric}}^{1,1}(\tilde{X})$, we recall that each point on $\Delta$ gives rise to a homogeneous coordinate.\footnote{We defined $\Delta^{\circ}$ as the dual (i.e., $N$-lattice) polytope for $X$, but are now studying the mirror $\tilde{X}$, for which $\Delta$ is the dual polytope.} Hence, na\"ively there are $\sum_{\text{codim}\Theta\geq 1}\ell^*(\Theta)$ toric divisors. However, points interior to facets correspond to ambient divisors that do not intersect the Calabi-Yau. Furthermore, there are in total four linear relations among the toric divisors. As a result, we obtain
\begin{equation}
h_{\text{toric}}^{1,1}(\tilde{X})=\sum_{\text{codim}\Theta\geq2}\ell^*(\Theta)-4\,.
\end{equation}
The Calabi-Yau $\tilde{X}$ considered in \S\ref{sec:example} has $h^{2,1}(\tilde{X})=3$ and $h^{1,1}(\tilde{X})=99$, and from the corresponding pair of dual polytopes $\Delta, \Delta^{\circ}$ one finds $h_{\text{toric}}^{2,1}(\tilde{X})=3$ and $h_{\text{toric}}^{1,1}(\tilde{X})=97$.  Thus all complex structure moduli of $\tilde{X}$ are toric, but two generators of the Picard group are non-toric. In order to determine the orientifold action on $H_4(\tilde{X},\mathbb{Z})$ we must therefore consider the non-toric divisors in more detail. Consider a point $\rho\in \Theta,$ with $\text{codim}\Theta=2$. Then $\{x_\rho=0\}\cap \tilde{X}$ is a reducible variety in $\tilde{X}$ and there are $1+\ell^*(\Theta^\circ)$ irreducible components \cite{Batyrev:1994hm,Braun:2017nhi}. Hence, there are in total $ \sum_{\text{codim}\Theta=2}\ell^*(\Theta)\ell^*(\Theta^\circ)$ non-toric divisors. As a result, we obtain
\begin{equation}
h^{1,1}(\tilde{X})=h_{\text{toric}}^{1,1}(\tilde{X})+\sum_{\text{codim}\Theta=2}\ell^*(\Theta)\ell^*(\Theta^\circ)\,.
\end{equation}
For a point $\rho,$ we call the divisor $\{x_\rho=0\}$ \emph{strictly favorable} if either $\rho$ is not interior to any two-face or it is interior to a two-face $\Theta$ but $\ell^*(\Theta^\circ)=0.$

Given isomorphisms $i^\circ:M\rightarrow \Bbb{Z}^4$ and $i:N\rightarrow \Bbb{Z}^4,$ we can assign coordinates to points in $\Delta^{\circ}\cap M$ and $\Delta\cap N.$  Including the origin, the points in $\Delta^{\circ}\cap \Bbb{Z}^4$ are
\begin{equation}
\Delta^{\circ}\cap \Bbb{Z}^4=\left(
\begin{array}{ccccccccc}
 0 & -1 & 1 & -1 & -1 & -1 & -1 & -1 & -1 \\
 0 & 3 & -1 & 0 & 0 & 0 & 0 & 0 & 1 \\
 0 & -2 & 0 & 0 & 0 & 1 & 2 & 1 & 0 \\
 0 & -1 & 0 & 1 & 0 & 1 & 0 & 0 & 0 \\
\end{array}
\right)\, ,
\end{equation}
where each column corresponds to a point.
The last point $(-1,1,0,0)$ is strictly interior to a facet.

We also record points of importance in the dual polytope $\Delta\cap\Bbb{Z}^4$:
\begin{equation}
(\Delta\cap \Bbb{Z}^4)\supset\left(
\begin{array}{ccccccccc}
 1 & 1 & -5 & -11 & 1 & 1&-5 & -4 & -3\\
 2 & 0 & -4 & -10 & 2 & 2&-4 & -4 & -3\\
 0 & 0 & 0 & -6 & 3 & 0 &-3 &-2 &-1\\
 0 & 0 & -6 & -6 & 0 & 6&0 &-3&-3\\
\end{array}
\right)\label{eqn:dual poly}
\end{equation}
The first six points in \eqref{eqn:dual poly} are the vertices of $\Delta\cap \Bbb{Z}^4.$  There is only one two-face $\check{\Theta} \in\Delta$ such that $\ell^*(\check{\Theta}^{\circ})\neq0$, and for this face $\ell^*(\check{\Theta}^{\circ})=1.$ The last two points in \eqref{eqn:dual poly} are strictly interior to $\check{\Theta}$, and hence each of those two points yields the union of two distinct divisors in $\tilde{X}$.
For notational simplicity we will suppress the dependence of the monomials on all the homogeneous coordinates except those that are explicitly presented in \eqref{eqn:dual poly}. We will denote by $x_i$ the coordinate given by the $i^{th}$ column.

The most general polynomial $f$ is
\begin{align}
f(\vec{x}) = &\psi_0x_1x_2x_3x_4x_5x_6x_7x_8 x_9-\psi_1x_1^6 -\psi_2 x_2^2x_8 x_9-\psi_3 x_4^6x_6^6x_7^6x_8^2x_9-\psi_4 x_3^6x_4^{12}x_7^6x_8^5x_9^4\nonumber\\ &-\psi_5 x_5^3x_6^6x_7^3-\psi_6 x_3^6x_5^6x_8 x_9^2-\psi_7 x_3^6x_4^6x_5^3x_7^3x_8^3x_9^3-\psi_8x_1^2x_3^2x_4^2x_5^2x_6^2x_7^2x_8 x_9\, .
\end{align}
The action of the root automorphism is
\begin{equation}\label{eq:root_auto}
x_2\mapsto x_2+\lambda x_1x_3x_4x_5x_6x_7\,.
\end{equation}
Hence, we confirm that $h^{2,1}(\tilde{X})=9-5-1=3.$

Now consider an orientifold action $\tilde{\mathcal{I}}: x_2\mapsto -x_2$.
The most general $\tilde{\mathcal{I}}$-invariant polynomial $f_{\tilde{\mathcal{I}}}$ contains 8 monomials,
\begin{align}
f_{\tilde{\mathcal{I}}}(\vec{x}) = &-\psi_1x_1^6 -\psi_2 x_2^2x_8 x_9-\psi_3 x_4^6x_6^6x_7^6x_8^2x_9-\psi_4 x_3^6x_4^{12}x_7^6x_8^5x_9^4\nonumber\\ &-\psi_5 x_5^3x_6^6x_7^3-\psi_6 x_3^6x_5^6x_8 x_9^2-\psi_7 x_3^6x_4^6x_5^3x_7^3x_8^3x_9^3-\psi_8x_1^2x_3^2x_4^2x_5^2x_6^2x_7^2x_8 x_9\, .
\end{align}
Because the nontrivial root automorphism of eq.~\eqref{eq:root_auto} does not commute with $\tilde{\mathcal{I}}$ we have $\text{dim}\,\mathcal{A}_{\tilde{\mathcal{I}}}=4$, where $\mathcal{A}_{\tilde{\mathcal{I}}}$ is the subgroup of $\mathcal{A}$ that commutes with the orientifold involution. As a result, there are in total three independent complex structure moduli, i.e.~$h_{-}^{2,1}(\tilde{X},\tilde{\mathcal{I}})=8-4-1=3.$

Next, we consider the structure of the non-toric divisors. To simplify the computation, we will blow down all of the blowup divisors whose blowup coordinates are implicit in eq.~\eqref{eqn:dual poly}.  First, we consider the locus $x_8=f=0$. We have
\begin{equation}
f|_{x_8=0}=-\psi_1 x_1^6-\psi_5x_5^3x_6^6x_7^3\,.\label{eqn:non favorable}
\end{equation}
We verified that $x_1=x_8=0$ does not intersect $X$ by confirming that the intersection numbers of $\{x_1=0\},$ $\{ x_8=0\}$, and $\{x_\rho=0\} $ are trivial for any $\rho$, i.e.~$x_1x_8$ is in the SR ideal. Hence, for $f|_{x_8=0}$ to have a solution, $x_5x_6x_7$ must not vanish. Using the toric rescaling, we set $\psi_1=-1$ and $\psi_5=1.$ Then na\"ively we obtain six disconnected solutions
\begin{equation}
x_1^2=\omega_3^i x_5x_6^2x_7\,,
\end{equation}
for $i=0,1,2$, where $\omega_3$ is a third root of unity. However, there is a $\Bbb{Z}_3$ subgroup of the Greene-Plesser group $G,$ \eqref{eqn:GP actions}, with the charge
\begin{equation}
\vec{\lambda}_{\Bbb{Z}_3}=(0,0,0,0,1,1,1)\, .
\end{equation}
This subgroup acts non-trivially on $x_5x_6^2x_7$,
\begin{equation}
\Bbb{Z}_3:x_5x_6^2x_7\mapsto \omega_3 x_5x_6^2x_7\,.
\end{equation}
One can verify that $G/\Bbb{Z}_3$ acts trivially on $(x_5x_6^2x_7)^{1/2}.$ As a result, there are two solutions to $x_8=f=0$:
\begin{equation}
x_1=\pm( x_5x_6^2 x_7)^{1/2} \,.\label{eqn:locus non fav}
\end{equation}
Likewise, the surface $x_9=f=0$ splits into the two solutions of eq.~\eqref{eqn:locus non fav}.

Using this we proceed to compute $h_-^{1,1}(\tilde{X},\tilde{\mathcal{I}})$.  Clearly, $\tilde{\mathcal{I}}$ acts trivially on the strictly favorable divisors $\{x_\rho=0\}$. Hence, we only need to check how $\tilde{\mathcal{I}}$ acts on the solutions of $x_8=f=0$ and $x_9=f=0.$ As \eqref{eqn:locus non fav} does not explicitly depend on $x_2$, the orientifold involution acts trivially on the solutions of \eqref{eqn:locus non fav}. Thus, the orientifold action on $H_4(\tilde{X},\mathbb{Z})$ is trivial, and so $h_-^{1,1}(\tilde{X},\tilde{\mathcal{I}})=0$.

Finally, we can compute the Euler characteristic $\chi(\mathfrak{F}_{\tilde{\mathcal{I}}})$ of the fixed locus $\mathfrak{F}_{\tilde{\mathcal{I}}}$ of $\tilde{\mathcal{I}}$ using the Lefschetz fixed point theorem,
\begin{equation}\label{q52}
-Q=\frac{\chi(\mathfrak{F}_{\tilde{\mathcal{I}}})}{4}=\frac{\chi(\tilde{X})}{4}+1+\Bigl(h_{-}^{2,1}(\tilde{X},\tilde{\mathcal{I}})-h_{-}^{1,1}(\tilde{X},\tilde{\mathcal{I}})\Bigr)=52 \,.
\end{equation}
This corresponds to the D3-brane tadpole.

\subsection{Computation in the orbifold limit}\label{App:2proveh11-=0}
The orientifold fixed locus in the orbifold $X/G$ contains the $G$-orbifold of the orientifold-fixed locus in $X$, but also further loci whose lifts in $X$ are the sets of points mapped by the orientifold to distinct points in the same $G$-orbit. It is straightforward to show that the full fixed locus $\mathfrak{F}_{\tilde{\mathcal{I}}}$ in $X/G$ is
\begin{equation}
\mathfrak{F}_{\tilde{\mathcal{I}}}\equiv\mathfrak{F}/G = (D_1\cup D_2\cup D_3 \cup D_4 \cup D_6)/G\, .
\end{equation}
We can compute the Euler characteristic of an orbifold as in \cite{Greene:1990ud,Candelas:1990rm}. We partition $\mathfrak{F}$ as $\mathfrak{F}=\cup_I \mathfrak{F}^f_{H_I}$ where the $\mathfrak{F}^f_{H_I}$ are the sets of points in $\mathfrak{F}$ that are fixed pointwise by subgroups $H_I\subset G$. Then, we have
\begin{equation}
\chi(\mathfrak{F}/G)=\sum_{I}\frac{\chi(\mathfrak{F}^f_{H_I})|H_I|}{|F_I|}\, ,
\end{equation}
where $F_I$ are the subgroups of $G/H_I$ that act freely (i.e. without fixed points) on $\mathfrak{F}^f_{H_I}$. As our group $G$ acts via multiplication by phases on the toric coordinates, the $\mathfrak{F}_{H_I}^f$ are the loci where a subset of the toric coordinates are set to zero, with subloci removed where further toric coordinates vanish. Such loci are mapped to themselves by all of $G$, so we have $F_I=G/H_I$. First, let us define toric divisors $D_i$, curves $\mathcal{C}_{ij}$ and sets of points $\mathcal{P}_{ijk}$ as
\begin{equation}
D_i=\{x_i=0\}\, ,\quad \mathcal{C}_{ij}=\{x_i=x_j=0\}\,  ,\quad \mathcal{P}_{ijk}=\{x_i=x_j=x_k=0\}\,  ,
\end{equation}
with pairwise distinct indices.
The $\mathfrak{F}_{H_I}^f$ can be chosen to be the above with lower-dimensional loci removed, i.e.
\begin{equation}
\hat{D}_i=D_i\backslash \Bigl(\cup_{j\neq i}\mathcal{C}_{ij}\bigcup \cup_{j\neq k\neq i}\mathcal{P}_{ijk}\Bigr)\, ,\quad \hat{C}_{ij}=\mathcal{C}_{ij}\backslash \cup_{k\neq i\neq j}\mathcal{P}_{ijk}\, ,\quad \hat{\mathcal{P}}_{ijk}=\mathcal{P}_{ijk}\, .
\end{equation}
Along the dense subset $\hat{D}_1\cup \hat{D}_2\cup \hat{D}_3 \cup \hat{D}_4 \cup \hat{D}_6$ in
$X$, the full group $G$ acts without fixed points. The Euler characteristics of the $D_i$
are $\chi_i=({45, 207, 11, 13, 24})$.

The curves $\hat{C}_{ij}$ are invariant under certain subgroups recorded on the left in Table \ref{tab:curves}.
\begin{table}
	\centering
	\begin{tabular}{c |c c c}
		$(i,j)$           &$|H|$ & $\chi$ & $\hat{\chi}$ \\ \hline	
		$(1,2)$           &$2$   & $-36$  & $-54$\\
		$(1,3)$           &$6$   & $0$    & $-6$\\
		$(1,4)$           &$6$   & $0$    & $-6$\\
		$(1,5)$           &$3$   & $-2$   & $-12$\\
		$(1,6)$           &$6$   & $-2$    & $-12$\\
		$(1,7)$           &$3$   & $4$     & $0$\\
		$(2,3)$           &$2$   & $-6$    & $-18$\\
		$(2,4)$           &$2$   & $-6$    & $-18$\\
		$(2,5)$           &$1$   & $-18$   & $-36$\\
		$(2,6)$           &$2$   & $-18$   & $-36$\\
		$(2,7)$           &$1$   & $12$    & $0$\\
		$(3,4)$           &$6$   & $0$     & $-6$\\
		$(3,5)$           &$3$   & $-2$    & $-12$\\
		$(4,5)$           &$3$   & $4$     & $0$\\
		$(4,7)$           &$3$   & $-2$    & $-12$\\
		$(5,6)$           &$3$   & $-2$    & $-12$\\
		$(6,7)$           &$3$   & $-2$    & $-12$\\
	\end{tabular}
	\begin{tabular}{c |c c }
		$(i,j,k)$           &$|H|$ & $\chi$  \\ \hline	
		$(1,2,3)$           &$12$   & $3$  \\
		$(1,2,4)$           &$12$   & $3$    \\
		$(1,2,5)$           &$6$   & $6$    \\
		$(1,2,6)$           &$12$   & $6$   \\
		$(1,2,7)$           &$6$   & $0$   \\
		$(1,3,4)$           &$36$   & $1$    \\
		$(1,3,5)$           &$18$   & $2$   \\
		$(1,4,5)$           &$18$   & $0$   \\
		$(1,4,7)$           &$18$   & $2$  \\
		$(1,5,6)$           &$18$   & $2$  \\
		$(1,5,7)$           &$6$   & $0$   \\
		$(1,6,7)$           &$18$   & $2$    \\
		$(2,3,4)$           &$12$   & $3$   \\
		$(2,3,5)$           &$6$   & $6$    \\
		$(2,4,5)$           &$6$   & $0$   \\
		$(2,4,7)$           &$6$   & $6$   \\
		$(2,5,6)$           &$6$   & $6$   \\
		$(2,5,7)$           &$6$   & $0$   \\
		$(2,6,7)$           &$6$   & $6$   \\
		$(3,4,5)$           &$18$   & $2$   \\
		$(4,5,7)$           &$18$   & $2$   \\
		$(5,6,7)$           &$18$   & $2$
	\end{tabular}
	\caption{Left: The curves $\mathcal{C}_{ij}$ invariant under subgroups $H_{ij}\subset G$, their Euler characteristics $\chi$, and the Euler characteristics $\hat{\chi}$ of the curves $\hat{\mathcal{C}}_{ij}$ obtained by removing toric points. Right: Analogous table for toric points.}
	\label{tab:curves}
\end{table}
The points $\mathcal{P}_{ijk}$ are invariant under the subgroups listed on the right in Table \ref{tab:curves}. The Euler characteristics of the non-compact curves $\hat{C}_{ij}$ are obtained from those of $\mathcal{C}_{ij}$ by subtracting the Euler characteristics of the points $\mathcal{P}_{ijk}\subset \mathcal{C}_{ij}$, which we also record in Table \ref{tab:curves}. The Euler characteristics $\hat{\chi}_i$ of the non-compact divisors $\hat{D}_i$, $i=1,2,3,5,6$ are likewise given by subtracting the Euler characteristics of the curves $\hat{C}_{ij}\subset D_i$ and points $P_{ijk}\subset D_i$. The result is
\begin{equation}
\hat{\chi}_i=({108, 324, 36, 36, 72})\, .
\end{equation}
Thus, finally, we obtain
\begin{align}
\chi(\mathfrak{F}_{\tilde{\mathcal{I}}})=\frac{1}{36}\sum_I \chi(\mathfrak{F}^f_{H_I})|H_I|^2=208\, ,
\end{align}
where the index $I$ collectively runs over the $\hat{D}_i$, $\hat{C}_{ij}$ and $\hat{P}_{ijk}$.

Thus, we confirm that
\begin{equation}
-Q=\frac{\chi(\mathfrak{F}_{\tilde{\mathcal{I}}})}{4}=52\, ,
\end{equation}
in agreement with eq.~\eqref{q52}.

\bibliography{refs}
\bibliographystyle{JHEP}
\end{document}